\documentclass{aastex63}

\received{Dec. 4, 2020}
\revised{Jan. 11, 2021}
\accepted{Feb. 15, 2021}
\submitjournal{PSJ}

\shorttitle{Jupiter dynamical tides}
\shortauthors{Idini \& Stevenson}
\graphicspath{{./}{figures/}}

\usepackage{bm}
\begin{document}

\title{Dynamical tides in Jupiter as revealed by Juno}

\correspondingauthor{Benjamin Idini}
\email{bidiniza@caltech.edu}

\author[0000-0002-2697-3893]{Benjamin Idini}
\affiliation{Division of Geological and Planetary Sciences, California Institute of Technology \\
1200 E California Blvd, MC 150-21 \\
Pasadena, CA 91125, USA}

\author[0000-0001-9432-7159]{David J. Stevenson}
\affiliation{Division of Geological and Planetary Sciences, California Institute of Technology \\
1200 E California Blvd, MC 150-21 \\
Pasadena, CA 91125, USA}



\begin{abstract}

    The Juno orbiter continues to collect data on Jupiter’s gravity field with unprecedented precision since 2016, recently reporting a non-hydrostatic component in the tidal response of the planet. At the mid-mission perijove 17, Juno registered a Love number \replaced{$k_2$}{$k_2=0.565\pm0.006$} that is \replaced{$-4.1\pm1.3\%$}{$-4\pm1\%$} ($1\sigma$) from the theoretical hydrostatic \replaced{$k_2$}{$k_2^{(hs)}=0.590$}. Here we assess whether the aforementioned departure of tides from hydrostatic equilibrium represents the neglected gravitational \replaced{effects}{contribution} of dynamical tides. We employ perturbation theory and simple tidal models to calculate a fractional dynamical correction $\Delta k_2$ to the well-known hydrostatic $k_2$. \added{Exploiting the analytical simplicity of a toy uniform-density model, we show how the Coriolis acceleration motivates the negative sign in the $\Delta k_2$ observed by Juno.} By simplifying Jupiter's interior into a core-less, fully-convective, and chemically-homogeneous body, we \replaced{derive two tidal models: a uniform-density sphere and an $n=1$ polytrope}{calculate $\Delta k_2$ in a model following an $n=1$ polytrope equation of state.} \deleted{Exploiting the analytical simplicity of the incompressible uniform-density model, we show how the Coriolis acceleration motivates the negative sign in $\Delta k_2$.} Our numerical results for the $n=1$ polytrope qualitatively follow the behaviour of the uniform-density model, mostly because the main component of the tidal flow is similar in each case. 
    Our results indicate that the gravitational effect of the Io-induced dynamical tide leads to $\Delta k_2=-4\pm1\%$, in agreement with the non-hydrostatic component reported by Juno. Consequently, our results suggest that Juno obtained the first unambiguous detection of the gravitational effect of dynamical tides in a gas giant planet. These results facilitate a future interpretation of Juno tidal gravity data with the purpose of elucidating the existence of a dilute core in Jupiter. 

\end{abstract}



\section{Introduction} \label{sec:intro}

The interior structure of a planet or star closely corresponds with its origin and evolution story. Seismology provides the tightest constraints on the interior structure of Earth (\cite{dahlen1998theoretical} and ref. therein), Saturn \citep{marley1993planetary,fuller2014saturn}, the Sun \citep{christensen1985seismology} and other distant stars (\cite{aerts2010asteroseismology} and ref. therein). In particular, Saturn’s ring seismology facilitates estimates of the planet’s rotation rate \citep{mankovich2019cassini} and possible dilute core (Mankovich et al., 2020). Unlike Saturn, Jupiter lacks extensive optically-thick rings with embedded waves that are excited by resonance of ring particle motions with internal normal modes. Alternatively to ring seismology, Doppler imaging reveals a suggested seismic behavior in Jupiter, limited to radial overtones of p-modes. At best, the current Doppler imaging data resolves the spacing in frequency space of low-order p-modes, providing a loose constraint compatible with simple interior models \citep{gaulme2011detection}. Future efforts based on similar techniques promise revealing additional information on Jupiter's seismic behavior. 

In the current absence of detailed seismological constraints, the Juno orbiter \citep{bolton2017jupiter} emerges as the alternative directed to reveal Jupiter's interior by employing gravity field measurements of global-scale motions. \replaced{Juno collects two kinds of gravity data that depend on properties of Jupiter's deep interior:}{Based on radiometric observations, Juno produces two kinds of gravity field measurements sensitive to Jupiter's interior structure:} the \deleted{even} zonal \replaced{$J_{2\ell}$}{$J_{\ell}$} and tesseral $C_{\ell,m}$ gravity coefficients. \added{The odd $J_{2\ell+1}$ coefficients reflect contributions from zonal flows, including atmospheric zonal winds and zonal flow in the dynamo region.} The \added{even} $J_{2\ell}$ coefficients contain Jupiter's response to the centrifugal effect responsible for Jupiter's oblateness, with \replaced{smaller}{minor} contributions from zonal winds in the atmosphere \citep{iess2018measurement} and the dynamo region \citep{kulowski2020contributions}. A time-dependent subset of $C_{\ell,m}$ coefficients contains Jupiter's tidal response to the gravitational pull from its system of satellites. Closely related to \added{the time-dependent} $C_{\ell,m}$, the Love number $k$ represent the non-dimensional gravitational field of tides evaluated at the outer boundary of the planet \citep{munk1960rotation}. One common interpretation relates $k$ to the degree of central concentration of the planetary mass (e.g., compressibility of the planetary material or the presence of a core). In a quadrupolar gravitational pull, the leading term in the tidal response relates to the Love number $k_2$\added{, which corresponds to the $\ell=m=2$ spherical harmonic}.

Following linear perturbation theory, the Love number $k_2$ breaks into two contributions: one hydrostatic and the other dynamic. The hydrostatic $k_2$ ignores the time–dependence of tides.
The dynamical tide represents the tidal flow and perturbed tidal bulge solving the traditional equation of motion $F^T=M\ddot{u}$, rather than $F^T=0$, where $F^T$ is the satellite-induced tidal force, diminished by the opposing self-gravity of the perturbed planet.

Early studies of the Love numbers in the gas giant planets relied on purely hydrostatic theory, assisted by simple thermodynamic principles alongside loosely constrained interior models \citep{gavrilov1977love}. With the assistance of historical astrometric data, the Cassini mission provided the first occasion to test the accuracy of those first hydrostatic models, finding a $\sim 10\%$ discrepancy between the theoretical and observed $k_2$ of Saturn \citep{lainey2017new}. Hydrostatic theory adjusted to Cassini's observation after incorporating the oblateness produced by the centrifugal effect into the tidal model\added{, not requiring to invoke the gravitational effects of Saturn dynamical tides}. The oblateness produced by the centrifugal effect is large due to Saturn's fast rotation, leading to major higher-order cross terms neglected in the earlier theory \citep{wahl2017concentric}. This effect remains hydrostatic provided that rotation occurs on cylinders, which allows the centrifugal force to be represented as the gradient of a potential. \deleted{The gravitational effects of Saturn dynamical tides remained opaque to Cassini because of the larger formal error in the observation.}

Unlike Cassini, the Juno orbiter recently detected a $3\sigma$ deviation in Jupiter's observed $k_2$ from the revised hydrostatic theory that accounts for the interaction of tides with oblateness \citep{notaro2019determination,durante2020jupiter}. Importantly, 
the difference between the observed $k_2$ and hydrostatic theory cannot be attributed to a failure to correctly constrain the hydrostatic number. Hydrostatic tides are well-constrained \added{(i.e., the $k_2$ error in \cite{wahl2020equilibrium} is $\pm0.02\%$ of the central value)} because the effect of oblateness of the planet on the zonal gravity coefficients $J_{2\ell}$ is known to a high precision by Juno. Juno's non-hydrostatic detection motivates a more careful consideration of neglected effects that contribute to $k_2$, particularly the gravitational field related to dynamical tides. 

Here we evaluate dynamical tides as a potential explanation to Juno’s non-hydrostatic detection. We concentrate on the contribution of dynamical tides to the overall gravity field while ignoring their contribution to dissipation. In other words, we implicitly assume that the imaginary part of $k_2$ is too small to affect our results for the real part, in agreement with observations of the orbital evolution of Io ($Q_2=-|k_2|/\textnormal{Im} (k_2) \sim 10^{5}$;  \cite{lainey2009strong}).

 We are motivated in our efforts by the prospect of finding an additional contribution to $k_2$ coming from Jupiter's core. A traditional model of a Jupiter-like planet consists of a discreet highly-concentrated central region of heavy elements (i.e., a core made of rock and ice) surrounded by a chemically homogeneous and adiabatic envelope of hydrogen-rich fluid material \citep{stevenson1982interiors}. Jupiter's observed radius indicates a super-solar abudance of heavy elements that accounts for a total of $\sim20M_E$. However, whether heavy elements reside in a traditional core or are distributed throughout the envelope is less clear. The observed $J_2$ and $J_4$ require some tendency towards central concentration but do not require a traditional core. In a subsequent investigation, we exploit our results of the dynamical tide presented here to answer questions about the origin and evolution of Jupiter by including in our model an enrichment of heavy elements that increases with depth (i.e., a dilute core; \cite{wahl2017comparing}). Using the information contained in $k_2$, we plan to provide answers to the following questions about Jupiter: Whether solid or fluid, does Jupiter have a traditional core? Alternatively, do the heavy elements in Jupiter spread out from the center forming a dilute core?

The remainder of this manuscript is organized as follows. In Section 2, we describe the Juno non-hydrostatic detection and develop the mathematical formalism used in the calculation of the fractional dynamical correction to $k_2$. In Section 3, we calculate the fractional dynamical correction to $k_2$ using simple tidal models, which leads to an unambiguous explanation to the non-hydrostatic Juno detection. In Section 4, we deliver a discussion on the limitations of our analysis, other physical processes potentially altering the Love number $k$, and future directions of investigation. In Section 5, we outline the conclusions and implications of our study.

\section{Jupiter's Love number}
\subsection{A correction to the hydrostatic $k_2$\label{sec:junok2}}
The main objective of this manuscript is to evaluate the hypothesis that Juno captured a systematic deviation of $k_2$ from the hydrostatic number and that most of the deviation can be explained by the neglected gravitational effect of dynamical tides.
The mean $k_2$ Juno estimate at the time of perijove 17 is $0.565$ \replaced{Notaro et al. 2019}{\citep{durante2020jupiter}}, establishing a \replaced{$-4.1\%$}{$-4\%$} deviation from the theoretical hydrostatic number $k_2^{(hs)} = 0.590$ \replaced{Wahl et al. (2016)}{\citep{wahl2020equilibrium}} \deleted{(Table~\ref{tab:love})}. 
The correction to the hydrostatic $k_2$ due to the rotational bulge is included in $k_2^{(hs)}$ and is of order $q = \Omega^2R_J^3/\mathcal{G}M_J \sim0.1$, the ratio of centrifugal effects to gravity at the equator \replaced{Wahl et al. (2016)}{\citep{wahl2020equilibrium}}; $R_J$ is the equatorial radius and $M_J$ Jupiter's mass.  
The \added{satellite-independent} $3\sigma$ uncertainty (confidence level $\approx 99.7\%$) in Juno's observation is 3$\%$ of $k_2^{(hs)}$ at perijove 17 \added{\citep{durante2020jupiter}}, close to the mean of observed deviation. At the end of the prime mission, the \added{satellite-independent} $3\sigma$ uncertainty in Juno's observation is projected to decrease to $ 1\%$ \added{(William Folkner, personal communication, April 8, 2020)}. Following the optimistic assumption that the mean deviation remains the same at the end of the prime mission, $k_2^{(hs)}$ will require a \added{non-hydrostatic} correction from $-5$ to $-3\%$ to be reconciled with $3\sigma$ observations. 

In our tidal models, we evaluate a dynamical correction to the hydrostatic $k_2$ as:
\begin{equation}
    \frac{k_2}{k_2^{(hs)}} = 1 + \Delta k_2
    + \mathcal{O}(1\%) \textnormal{,}
    \label{eq2:corrections}
\end{equation}
 where $\Delta k_2$ is the fractional dynamical correction calculated for a spherical planet using perturbation theory and comes from the inertia terms in the equation of motion. 
 The fractional dynamical correction is of order $\Delta k_2\sim\omega^2/4\pi\mathcal{G}\bar{\rho}\sim0.1$, where $\bar{\rho}$ is the mean density of the planet and $\omega$ the forcing frequency related to the tide. The theoretical hydrostatic number $k_2^{(hs)}$ includes the effect of the oblateness of the planet, a realistic equation of state, and a density profile consistent with the zonal gravitational moments $J_2$ and $J_4$. From assuming that the number $k_2^{(hs)}$ is perfectly known, we aim to evaluate how the gravitational \replaced{effect}{contribution} of dynamical tides perturbs $k_2^{(hs)}$. 
 
\begin{figure}[ht!]
    \plotone{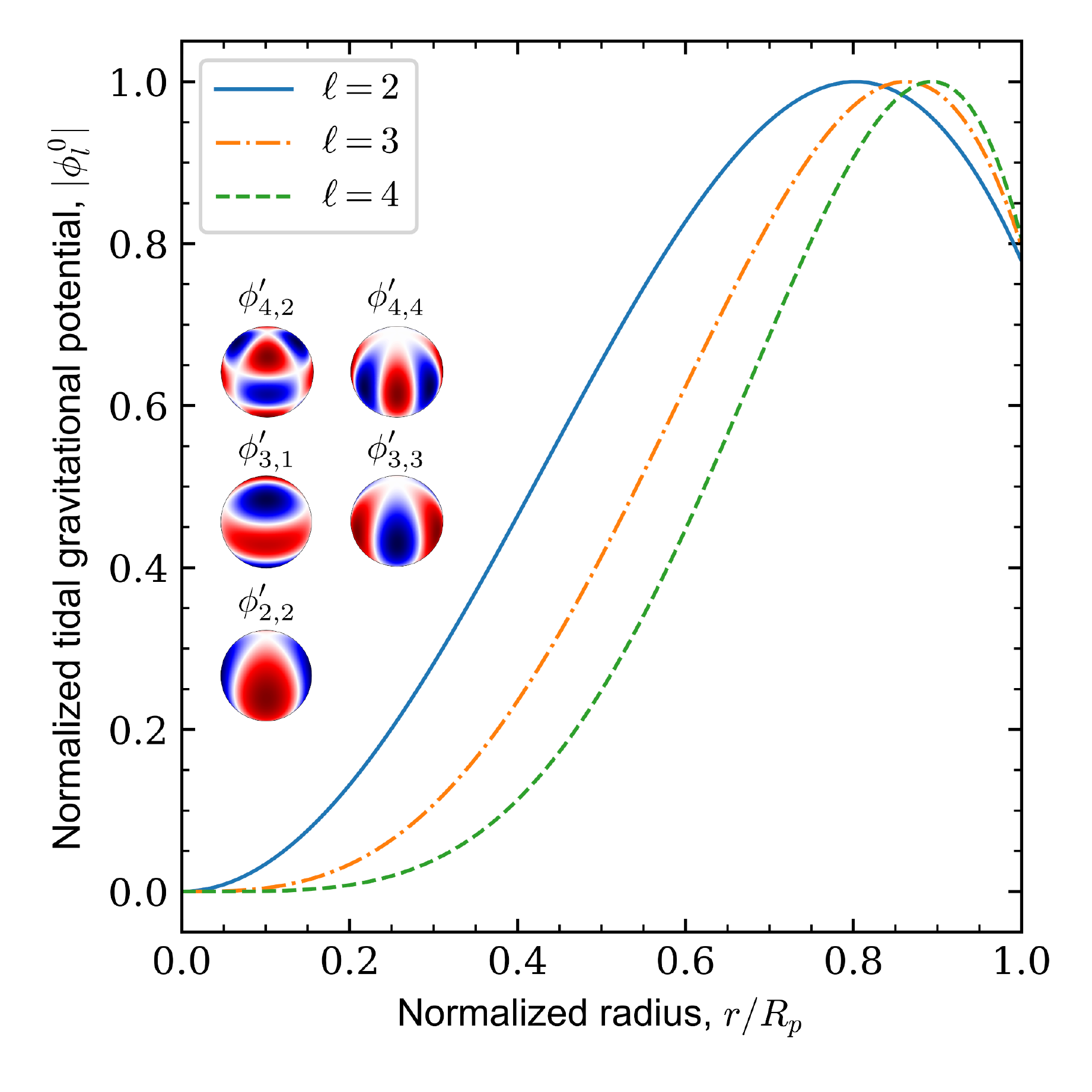}
    \caption{Angular patterns and radial functions describing the 3-D structure of the gravitational field of tides. The radial functions are the normalized hydrostatic gravitational potential $|\phi^0|$ in an $n=1$ polytrope (\ref{eqa:phi0}). The sign of the hydrostatic $\ell-$tide is $(-1)^{\ell/2 +1}$. \label{fig:hsradial}}
\end{figure}

 Instead of adding dynamical effects into the already complicated \added{numerical} model used to calculate $k_2^{(hs)}$, we use perturbation theory to isolate the dynamical effects in a much simpler interior model defined by an $n=1$ polytropic equation of state. The $n=1$ polytrope $p=K\rho^2$ closely follows the equation of state of a H-He mixture (Stevenson 2020) and is chosen for computational simplicity but is not crucial to the \added{$\Delta k_2$} calculation. \added{The density distribution in a non-rotating $n=1$ polytrope is $\rho = \rho_c j_0 (kr)$. The central density $\rho_c$ is set to fit Jupiter's total mass; $j_0$ is the zero-order spherical Bessel function of second kind; $k^2=2\pi\mathcal{G}/K$ is a normalizing constant for the radius, where $K = 2.1\cdot10^{12}$ (cgs) for a H/He cosmic ratio; $\mathcal{G}$ is the gravitational constant; $r$ is the radial coordinate. Exploiting the compact equation of state and density profile of the $n=1$ polytrope, we calculate the dynamical Love number by accounting for the dynamical terms in the equation of motion (Section~\ref{sec:n1pol}). The fractional dynamical correction $\Delta k_2$ comes from comparing the hydrostatic and dynamical Love number in the polytrope; however, as the correction is expressed in fractional terms, $\Delta k_2$ calculated this way introduces dynamical effects into any hydrostatic model, to leading order approximation. } The hydrostatic Love number in a spherical planet following an $n=1$ polytrope is obtained analytically (Fig.~\ref{fig:hsradial}; Appendix~\ref{sec:hs}). For example, the degree-2 hydrostatic Love number of a spherical planet following an $n=1$ polytrope is $k_2 = 15/\pi^2 - 1\approx 0.520$.  \deleted{The density distribution in a non-rotating $n=1$ polytrope is $\rho = \rho_c j_0 (kr)$. The central density $\rho_c$ is set to fit Jupiter's total mass; $j_0$ is the zero-order spherical Bessel function of second kind; $k^2=2\pi\mathcal{G}/K$ is a normalizing constant for the radius, where $K = 2.1\cdot10^{12}$ (cgs) for a H/He cosmic ratio; $\mathcal{G}$ is the gravitational constant; $r$ is the radial coordinate. Exploiting the compact equation of state and density profile of the $n=1$ polytrope, we calculate the dynamical Love number by accounting for the dynamical terms in the equation of motion. The fractional change to the hydrostatic $k_2$ comes from these two results in the polytrope; however, as the correction is express in fractional terms, $\Delta k_2$ calculated this way may apply to any hydrostatic model, to leading order approximation.} Stated explicitly, our approach assumes that the realistic elements included in the calculation of $k_2^{(hs)}=0.590$ \replaced{mostly matter for the hydrostatic $k_2$ and their}{produce little} effect on the dynamical tide \added{and} can be ignored for now.

The uncertainty $\mathcal{O}(1\%)$ in equation (\ref{eq2:corrections}) is an order-of-magnitude estimate of the neglected cross term that accounts for the effect of the centrifugal effect on dynamical tides. Individually, the centrifugal effect and the dynamical effect are both of order $\sim 10\%$. In linear perturbation theory, the crossed term is roughly the multiplication of the individual terms, resulting in $\sim1\%$. The uncertainty may be smaller as the dynamical correction involves terms that tend to cancel each other. This cancellation is expected to remain when account is made of the oblateness of the planet, since (as our analysis shows below) it is the small difference between $\omega$ and $2\Omega$ that matters, with or without oblateness.
 
The \replaced{reminder}{remainder} of this manuscript deals with the calculation of the dynamical Love number.

\subsection{Equations of tides in an adiabatic gas giant planet}

To calculate the Love number $k$, we require to compute the tidal gravitational potential $\phi'$ on Jupiter. The Love number $k$ represents the ratio of $\phi'$ over the gravitational pull of the satellite $\phi^T$ evaluated at the outer boundary of the planet:
\begin{equation}
    k_{\ell,m}= \left(\frac{\phi'_{\ell,m}}{\phi^T_{\ell,m}}\right)_{r=R_p} \textnormal{.}
\end{equation}
At a distance $r$ from the center of a planet of radius $R_p$, the potential $\phi^T$ from a satellite orbiting in a circular orbit aligned with the equatorial plane of the planet is:
\begin{equation}
    \phi^T = \sum_{\ell=2}\sum_{m=-\ell}^\ell U_{\ell,m} \left(\frac{r}{R_p}\right)^\ell  Y_\ell^m(\theta,\varphi)e^{i\omega t} \textnormal{,}
    \label{eq1:forcing}
\end{equation}
where $U_{\ell,m}$ are a numerical constants and $Y_\ell^m$ normalized spherical harmonics (Appendix~\ref{sec:hs}). The tidal frequency $\omega = |m(\Omega-\omega_s)|$ represents the frequency of a standing wave as observed from the perspective of an observer rotating with the planet at spin rate $\Omega$, where $\omega_s$ is the orbital frequency of the satellite and $m$ the order of the tide.  We follow the convention where $\omega$ is always positive and retrograde tides are represented by a negative order $m$ that flips the coordinate frame. The simplifications applied to the orbit are consistent to first order with the observed eccentricities $e<0.01$ and inclinations $i<0.5^\circ$ of the Galilean satellites.

We calculate the tidal response of a rigidly-rotating planet from a problem defined by the linearly-perturbed momentum, continuity, and Poisson's equations, respectively:
\begin{equation}
     -i\omega \bm{v} + 2\bm{\Omega} \times \bm{v} = - \frac{\nabla p'}{\rho} + \frac{\rho'}{\rho^2}\nabla p + \nabla\tilde{\phi}'  \textnormal{,}
     \label{eq1:momentum}
\end{equation}
\begin{equation}
     \nabla\cdot(\rho \bm{v}) = i\omega\rho'\textnormal{,}
     \label{eq1:continuity}
\end{equation}
\begin{equation}
    \nabla^2\tilde{\phi}' = -4\pi\mathcal{G}\rho'\textnormal{.}
    \label{eq1:poisson}
\end{equation}

The tidal response of the planet produces adiabatic perturbations to the gravitational potential $\phi'$, the density profile $\rho'$, and pressure $p'$. The potential of the gravitational pull $\phi^T$ and the tidal gravitational potential $\phi'$ are combined into $\tilde{\phi}'=  \phi^T+\phi'$ for analytical simplicity. Adiabatic perturbations in an adiabatic planet follow the thermodynamic statement (e.g., \cite{wu2005origin}):
\begin{equation}
    \frac{p'}{ p} = \Gamma_1\frac{\rho'}{\rho} = c_s^2\frac{\rho'}{p}  \textnormal{,}
\end{equation}
where $\Gamma_1$ is the first adiabatic index \citep{aerts2010asteroseismology}. Unprimed pressure and density represent the unperturbed state of the planet in hydrostatic equilibrium. We rewrite the momentum equation of an adiabatic planet as:

\begin{equation}
     -i\omega \bm{v} + 2\bm{\Omega} \times \bm{v} =\nabla \left( \tilde{\phi}' - \frac{p'}{\rho}\right) =\nabla \left( \tilde{\phi}' - c_s^2\frac{\rho'}{\rho}\right) =\nabla\psi  \textnormal{.}
     \label{eq1:momentum2}
\end{equation}
In equation (\ref{eq1:momentum2}), hydrostatic tides follow $\psi=0$ (Appendix~\ref{sec:hs}).

The tidal flow becomes a function of the potential $\psi$ after operating the divergence and curl on the momentum equation \citep{wu2005origin,goodman2009dynamical}:
\begin{equation}
    \bm{v} = -\frac{i\omega }{4\Omega^2-\omega^2}\left(\nabla\psi + \frac{2}{i\omega} \mathbf{\Omega}\times\nabla\psi - \frac{4}{\omega^2}\mathbf{\Omega}(\mathbf{\Omega}\cdot\nabla\psi) \right)\textnormal{.}
    \label{eq2:v}
\end{equation}

After replacing the flow into the continuity equation, the governing equations of tides in an adiabatic planet reduce to:
\begin{equation}
    \nabla \cdot \left(\rho\left(\nabla\psi + \frac{2}{i\omega} \mathbf{\Omega}\times\nabla\psi - \frac{4}{\omega^2}\mathbf{\Omega}(\mathbf{\Omega}\cdot\nabla\psi) \right)\right) =  \left(\frac{4\Omega^2 - \omega^2}{4\pi\mathcal{G}}\right) \nabla^2\tilde{\phi}'  
    \label{eq2:psi}\textnormal{,}
\end{equation}
\begin{equation}
    \psi = \frac{c_s^2}{4\pi\mathcal{G}\rho}\nabla^2\tilde{\phi}'+\tilde{\phi}'\textnormal{.}
    \label{eq2:phi}
\end{equation}

Perturbation theory allows us to decouple the weakly coupled potentials $\psi$ and $\tilde{\phi}'$ in equations (\ref{eq2:psi}) and (\ref{eq2:phi}).
According to perturbation theory, the tidal gravitational potential splits into a static and dynamic part:
 \begin{equation}
     \phi' = \phi^0 + \phi^{dyn} \textnormal{,}
 \end{equation}
 where $\phi^0$ corresponds to the gravitational potential of the hydrostatic tide after solving equation (\ref{eq2:phi}) with $\psi =0$ (Fig.~\ref{fig:hsradial}; Appendix~\ref{sec:hs}). By definition, the sound speed in an $n=1$ polytrope follows $c^2_s = 2K \rho$, which reduces the hydrostatic version of equation (\ref{eq2:phi}) to:
\begin{equation}
\label{eq:8}
    \frac{\nabla^2\phi^0}{k^2} + \phi^0+\phi^T = 0\textnormal{.}
\end{equation} 
 As a good approximation, we ignore the contribution from dynamical tides to the potential $\psi$ by setting $\nabla^2\tilde{\phi}'\approx \nabla^2\phi^0$ in the right-hand side in equation (\ref{eq2:psi}). According to equation (\ref{eq2:psi}), the potentials satisfy $\psi \sim  \phi'\omega^2/4\pi\mathcal{G}\bar{\rho}$, which leads to $\phi'\gg\psi$ given a tidal frequency $\omega^2\ll 4\pi\mathcal{G}\bar{\rho}$. By continuity, the approximation means that the tidal flow mostly advects the mass in the hydrostatic tidal bulge (i.e., $\rho'\approx\rho^0$ and equation (\ref{eq1:poisson})). The decoupled tidal equations simply to:
 \begin{equation}
    \nabla \cdot \left(j_0(kr)\left(\nabla\psi + \frac{2}{i\omega} \mathbf{\Omega}\times\nabla\psi - \frac{4}{\omega^2}\mathbf{\Omega}(\mathbf{\Omega}\cdot\nabla\psi) \right)\right) =  \left(\frac{4\Omega^2 - \omega^2}{4\pi\mathcal{G}\rho_c}\right) \nabla^2\phi^0  \textnormal{,}
    \label{eq2:psi_n1}
\end{equation}
\begin{equation}
    \psi = \frac{\nabla^2\phi^{dyn}}{k^2}+\phi^{dyn}\textnormal{.}
    \label{eq2:phi_n1}
\end{equation}
 We obtain the dynamical gravitational potential $\phi^{dyn}$ first solving $\psi$ from equation (\ref{eq2:psi_n1}) and then using the result to calculate $\phi^{dyn}$ from equation (\ref{eq2:phi_n1}). 

The boundary condition at the center of the planet imposes a finite solution for both potentials, allowing us to discard the divergent term characteristic of problems that include the Laplace operator. As required for a free planetary boundary, the condition at the outer boundary sets the Lagrangian perturbation of pressure equal to zero (e.g., \cite{goodman2009dynamical}):
\begin{equation}
    v_r=\bm{v}\cdot\hat{n} = -i\omega\frac{p'}{\partial_r p}
    =-i\omega\rho'\frac{ c_s^2}{\rho g} \textnormal{,}
\end{equation}
or:
\begin{eqnarray}
   \hat{n}\cdot\nabla\psi +  \frac{2}{i\omega}  \hat{n}\cdot(\mathbf{\Omega}\times\nabla\psi) - \frac{4}{\omega^2}( \hat{n}\cdot\mathbf{\Omega})(\mathbf{\Omega}\cdot\nabla\psi)&=& -\left(\frac{4\Omega^2-\omega^2}{4\pi\mathcal{G}}\right)\frac{ c_s^2}{ \rho g} \nabla^2\tilde{\phi}'\nonumber\\
   &=&-\left(\frac{4\Omega^2-\omega^2}{g}\right)(\psi-\tilde{\phi}')\textnormal{,}
   \label{eq2:bcout}
\end{eqnarray}
where $v_r$ is the radial component of the tidal flow, $\hat{n}$ is a unitary vector normal to the outer boundary, and $g$ the gravitational acceleration at the outer boundary. The sound speed and density nearly vanish near the outer boundary of a compressible body, resulting in a finite radial flow (e.g., $v_r = -2K\omega\rho'/g$ in an $n=1$ polytrope). The outer boundary condition (\ref{eq2:bcout}) indicates $\psi/R_J \sim (\tilde{\phi}'-\psi)\omega^2/g$. The Jupiter-Io system prescribes $1/R_J\gg\omega^2/g$, which leads to $\tilde{\phi}'\gg \psi$ and simplifies equation (\ref{eq2:bcout}) into:
\begin{equation}
   \hat{n}\cdot\nabla\psi +  \frac{2}{i\omega}  \hat{n}\cdot(\mathbf{\Omega}\times\nabla\psi) - \frac{4}{\omega^2}( \hat{n}\cdot\mathbf{\Omega})(\mathbf{\Omega}\cdot\nabla\psi)=\left(\frac{4\Omega^2-\omega^2}{g}\right)(\phi^0+\phi^T)\textnormal{.}
   \label{eq2:bcout2}
\end{equation}
Also at the outer boundary, the gravitational potential should be continuous in amplitude and gradient with a gravitational potential external to the planet that decays as $r^{-(l+1)}$.

In the following section, we solve the tidal equations for a uniform-density model and an $n=1$ polytrope model, with and without the Coriolis effect.

\section{Dynamical tides in a gas giant planet}
 Following our simplified model of dynamical tides, we calculate $\Delta k_2$ in a coreless, chemically homogeneous, and adiabatic Jupiter-like model. The thermal state becomes almost adiabatic in a convecting fluid planet with homogeneous composition. From the point of view of tidal calculations, the deviation from adiabaticity is negligible in the interior because the superadiabaticity required to sustain convection is a tiny fraction of the adiabatic temperature gradient, despite the possible inhibitions arising from rotation and convection. A fluid parcel in an adiabatic interior that is adiabatically displaced by a tidal perturbation will find itself at a new state that is essentially unchanged in density and temperature from the unperturbed state at that pressure. This definition of neutral stability begins to break down near the photosphere, where the density is low and the radiative time constant is no longer huge for blobs with spatial dimension of order the scale height. However, that region represents only a tiny fraction of the planet and does not produce enough gravity to significantly alter the real part of the Love number $k$. We discuss hypothetical contributions to $k$ from a core and depth-varying chemical composition in Section 4.

As a matter of simplifying the arguments presented in this section, we mostly concentrate on the Love number at $\ell=m=2$, commonly known as $k_2$. Correspondingly, $k_2$ is forced by the degree-2 component in the gravitational pull: 
\begin{equation}
    \phi^T_2 = \frac{3}{16}\frac{\mathcal{G}m_s}{a^3} r^{2}\sin^2\theta  e^{-i(\omega t+2\varphi)} \textnormal{.}
    \label{eq1:forcing2}
\end{equation}
\added{Dynamical effects scale with the satellite-dependent $\omega$. We concentrate in the dynamical effects caused by Io, the Galilean satellite with the dominant gravitational pull on Jupiter.} 

\subsection{A non-rotating gas giant\label{sec:norot}}
To an excellent approximation, dynamical tides in a non-rotating planet represent the forced response of the planet in the fundamental normal mode of oscillation (f-mode) \citep{vorontsov1984dynamical}. Despite Cassini suggesting that higher-order normal mode overtones (p-modes) dominate the gravitational field of Saturn's free-oscillating normal modes \citep{markham2020possible}, the forced response of normal modes depends on the coupling of the gravitational-pull and the mode radial eigenfunction. The forced response of p-modes contribute negligibly to the gravitational field of dynamical tides \citep{vorontsov1984dynamical} because of the bad coupling between the zero-node radial component of the gravitational pull and p-modes' eigenfunctions, the later having one or more radial nodes. Conversely, the gravitational pull more efficiently excites f-modes, whose radial eigenfunctions roughly follow the radial scaling of the gravitational pull ($\propto r^\ell$).

\subsubsection{The harmonic oscillator analogy}
In the following, we use the forced harmonic oscillator as an analog model to tidally-forced f-modes. In this model, the fractional dynamical correction to $k_2$ acquires a simple analytical form. The equation of motion of a mass $M$ connected in harmonic motion to a spring of stiffness $\mathcal{K}$ and negligible dissipation is:
\begin{equation}
    -M  \omega^2 u+ \mathcal{K}u = F^T \textnormal{,}
\end{equation}
where $\omega$ is the forcing frequency and $F^T$ the tidal forcing.  F-modes oscillate at frequencies $\omega_0$ that are much higher than the forcing tidal frequency, meaning that tidal resonances with f-modes are highly unlikely. Assuming that dynamical effects are small so that the tidal forcing is mostly balanced by static effects (i.e., $F^T \approx \mathcal{K} u_s$), the displacement of the mass is:
\begin{equation}
u = u_s\left( \frac{\omega^2_0}{\omega^2_0-\omega^2}\right)\textnormal{.}
\end{equation}
The mass assumes the static equilibrium position $u_s$ as the forcing frequency tends to zero. The displacement $u$ is analogous to the Love number $k$, thus the fractional dynamical correction becomes:
\begin{equation}
    \Delta k = \frac{u-u_s}{u_s}=\frac{\omega^2}{\omega^2_0-\omega^2}\textnormal{.}
    \label{eq2:dk2}
\end{equation}

\subsubsection{The Coriolis-free $n=1$ polytrope \label{sec:norot2}}
To verify the analogy of the forced harmonic oscillator to tidally-forced f-modes, we calculate the tidal response of a non-rotating $n=1$ polytrope directly from the governing equations of tides. When $\Omega =0$, the governing equation (\ref{eq2:psi_n1}) reduces to:
 \begin{equation}
     j_0(kr)\nabla^2\psi-j_1(kr)\partial_r(\psi) =-\left(\frac{\omega^2}{4\pi\mathcal{G}\rho_c}\right)\nabla^2\phi^0 \textnormal{.}
      \label{eq3:approximation}
 \end{equation}

For the potential $\psi$ at $\ell=m=2$, the boundary condition at the outer boundary $r=R_p$ (\ref{eq2:bcout}) is:
\begin{equation}
    \frac{g}{\omega^2}\partial_r\psi_{2} -\psi_{2} = -5j_2(kR_p) \textnormal{.}
\end{equation}
For the same degree and order, the continuity of the gravitational potential and its gradient at the outer boundary requires:
\begin{equation}
    \partial_r\phi^{dyn}_2  = -\left(\frac{3\phi^{dyn}_2 + 5}{R_p}\right)\textnormal{.}
\end{equation}
At the center of the planet $r= r_0\to0$, we find the following scaling: $\nabla^2\phi^0\sim0$, $j_1(kr_0)\sim0$, and $j_0(kr_0)\sim \textnormal{constant}$. A finite potential $\psi$ satisfying equation (\ref{eq3:approximation}) is $\psi_2\sim r^2$ near $r_0$, or:
\begin{equation}
    \partial_r\psi_2 -\frac{2}{r_0}\psi_{2} =0\textnormal{.}
\end{equation}
Similarly, a finite gravitational potential of dynamical tides is $\phi^{dyn}_{2}\sim r^2$ at the center of the planet, satisfying:
\begin{equation}
    \partial_r\phi^{dyn}_2 -\frac{2}{r_0}\phi^{dyn}_{2} =0\textnormal{.}
\end{equation}

We compute the fractional dynamical correction to $k_2$ first projecting the tidal equations into spherical harmonics (Appendix~\ref{sec:sh}) and later solving for the relevant potentials using a Chebyshev pseudo-spectral numerical method (Appendix~\ref{sec:cheb}). After projecting equations (\ref{eq3:approximation}) and (\ref{eq2:phi_n1}) into spherical harmonics, we obtain two decoupled equations for the radial parts of the potentials $\psi$ and $\phi$ (Appendix~\ref{sec:sh1}).
After numerically solving the radial equations in Appendix~\ref{sec:sh1} using Io's gravitational pull \replaced{($\omega=2\omega_s\approx 81$ $\mu$Hz)}{$\omega_s\approx 42\mu$Hz)}, the fractional dynamical correction corresponds to $\Delta k_2\approx 1.2\%$, in close agreement with the forced harmonic oscillator analogy applied to the oscillation frequency of the degree-2 f-mode $\omega_0 \approx 740$ $\mu$Hz \citep{vorontsov1976free}. \added{We observe a similar agreement  between the harmonic oscillator and the Coriolis-free $n=1$ polytrope at higher-degree spherical harmonics (Table~\ref{tab:love0}).}
Our results agree with a previously reported fractional correction to the gravitational coefficient $C_{2,2}\propto k_2$ due to dynamical tides in a non-rotating Jupiter. \citep{vorontsov1984dynamical}. 

\begin{deluxetable*}{ccc}
\tablenum{1}
\tablecaption{Io-induced fractional dynamical correction $\Delta k$ in a Coriolis-free Jupiter.\label{tab:love0}}
\tablewidth{0pt}
\tablehead{
\colhead{} & \colhead{Harmonic oscillator} & \colhead{$n=1$ polytrope} \\
\colhead{Type} & \colhead{($\%$)} & \colhead{($\%$)}
}
\decimalcolnumbers
\startdata
$\Delta k_2$    & +15 & +13  \\
$\Delta k_{42}$  & +5& +5 \\
$\Delta k_{31}$  & +2 & +2 \\
$\Delta k_{33}$  & +19& +15 \\
$\Delta k_{44}$ & +25& +19 \\
\enddata
\tablecomments{(2) See equation (\ref{eq2:dk2}). The mode frequency without rotation comes from \cite{vorontsov1976free}.}
\end{deluxetable*}

When \cite{vorontsov1984dynamical} excluded Jupiter’s spin, they were doing something that was mathematically sensible but physically peculiar: tides occur much more frequently in Jupiter’s rotating frame of reference and the tidal flow is accordingly much larger than if you had Jupiter at rest, which implies a much larger dynamical effect. Consequently, $\Delta k_2$ increases by an order of magnitude after partially including Jupiter's rotation in the tidally-forced response of f-modes. Without the Coriolis effect but including Jupiter's spin rate ($\Omega \approx 176$ $\mu$Hz) in the calculation of Io's tidal frequency ($\omega\approx 270$ $\mu$Hz), the fractional dynamical correction in an $n=1$ polytrope corresponds to $\Delta k_2 \approx 13\%$, close to the $\Delta k_2 \approx 15\%$ from the forced harmonic oscillator analogy (\ref{eq2:dk2}). In general for a non-rotating planet, the dynamical correction increases as the tidal frequency approaches the characteristic frequency of Jupiter's f-modes ($\sqrt{\mathcal{G}M_J/R_J^3}\sim 600$ $\mu$Hz).

\subsection{The Coriolis effect in a rotating gas giant}
The Galilean satellites produce dynamical tides for which the Coriolis effect plays an important role.
Following relatively slow orbits ($\omega_s\ll \Omega$), the Galilean satellites produce tides on Jupiter with a tidal frequency  $\omega\sim 2\Omega$. 
Consequently, the two inertial terms responsible for dynamical tides in the left-hand side in the equation of motion (\ref{eq1:momentum}) have similar amplitude. Moreover, Juno observes $k_2$ to be less than the predicted number for a purely hydrostatic tide (Section~\ref{sec:junok2}) and yet our analysis above produces a positive $\Delta k_2$ when dynamical effects are included and the Coriolis effect neglected (see equation (\ref{eq2:dk2})). We must accordingly motivate the change in sign when Coriolis is included. 

In the following, we first calculate the gravitational effect of dynamical tides in a uniform-density sphere to reveal the fundamental behaviour of the tidal equations avoiding most of the technical difficulties related to using an $n=1$ polytrope. We later found that the more complicated case of an $n=1$ polytrope introduces a minor quantitative difference, but lead to the same general behaviour. 

\subsubsection{A uniform-density sphere}
First, we explain why $\Delta k_2$ changes sign because of the Coriolis effect in a specially simple model with uniform density. We calculate the fractional dynamical correction to $k_2$ in two steps: (1) we calculate the potential of the flow $\psi$ in a uniform-density sphere, and (2) we use the $\psi$ calculated this way to calculate the gravity potential $\phi'$. 

In a uniform-density sphere, the sound speed $c_s$ is infinite and (\ref{eq2:psi}) reduces to the well-known Poincare problem \citep{greenspan1968theory}:
\begin{equation}
    \nabla^2\psi - \frac{4}{\omega^2}(\bm{\Omega}\cdot\nabla)^2\psi = 0 \textnormal{,}
    \label{eq3:uniform}
\end{equation}
where the boundary condition at the outer boundary requires to satisfy equation (\ref{eq2:bcout2}).

Following the \replaced{incompresibility}{incompressibility} of a uniform-density sphere, $\psi_2$ retains the symmetry and degree-2 angular structure from the gravitational pull in equation (\ref{eq1:forcing2}), thus acquiring exact solutions in the form:
\begin{equation}
    \psi_2 \propto (x-iy)^2\textnormal{.}
\end{equation}
The numerical factor in $\psi_2$ is set by the outer boundary condition (\ref{eq2:bcout}), corresponding to \citep{goodman2009dynamical}:
\begin{equation}
    \psi_2 = \frac{3\omega(2\Omega-\omega)}{8\pi\mathcal{G}\bar{\rho}} \tilde{\phi}_2'= \frac{R_p\omega(2\Omega-\omega)}{2g} \tilde{\phi}_2'\textnormal{.}
    \label{eq3:psi2}
\end{equation}

In a constant-density sphere, tides act displacing the sphere's boundary within an infinitesimally thin shell. According to the momentum equation, the tidal gravitational potential relates to the potential $\psi$ following:
\begin{equation}
    \tilde{\phi}' - \psi -\frac{p'}{\bar{\rho}} =0\textnormal{.}
    \label{eq:unik2}
\end{equation}
The radial tidal displacement projected into spherical harmonics is:
\begin{equation}
    \xi = \sum_{\ell,m}\xi_{\ell,m}(r)Y_\ell^m(\theta,\varphi)\textnormal{,}
\end{equation}
and the pressure perturbation follows:
\begin{equation}
    p' = -\xi \frac{\partial p}{\partial r} = \xi\bar{\rho} g\textnormal{.}
\end{equation}
The gravitational potential of a thin spherical density perturbation follows directly from the definition of the gravitational potential and integration throughout the volume:
\begin{equation}
    \phi' = \sum_{\ell,m} \frac{4\pi \mathcal{G}\bar{\rho}}{(2\ell +1)}\frac{R_p^{\ell+2}}{r^{\ell+1}}\xi_{\ell,m}Y_\ell^m\textnormal{.}
\end{equation}
The degree-2 tidal gravitational potential corresponds to:
\begin{equation}
    \phi_2'= \frac{4\pi\mathcal{G}\bar{\rho}}{5}\frac{R_p^4}{r^3}\xi_2 = \frac{3}{5}\left(\frac{R_p}{r}\right)^3g\xi_2\textnormal{.}
\end{equation}
We once again use perturbation theory to split the hydrostatic and dynamic contributions to the tidal displacement (i.e., $\xi_2 = \xi^0+\xi^{dyn} $). We first solve the well-known problem of the hydrostatic $k_2$ (i.e., $\psi = 0$) in a uniform-density sphere \citep{love1909yielding}. At the sphere's boundary (i.e., $r=R_p$), the hydrostatic gravitational potential follows $\phi^0 = 3g\xi^0/5$. From equation (\ref{eq:unik2}) evaluated at $r=R_p$, the potential of the gravitational pull becomes $\phi^T=2g\xi^0/5$. Following the last two results, the Love number is $k_2=3/2$, as expected. 

The dynamical contribution to the tidal displacement $\xi^{dyn}$ produces the gravitational potential $\phi^{dyn}=3g\xi^{dyn}/5$. After applying perturbation theory and cancelling the hydrostatic terms in equation (\ref{eq:unik2}), the potential $\psi_2$ becomes $\psi_2 = -2g\xi^{dyn}/5$. Combined with equation (\ref{eq3:psi2}), the last result for $\psi_2$ allows us to reach an expression for the fractional dynamical correction in a uniform-density sphere:
\begin{equation}
 \Delta k_2 =  \frac{\xi^{dyn}}{\xi^0}  \approx - \left(\frac{5R_p}{4g}\right)\omega(2\Omega-\omega)\textnormal{.}
    \label{eq3:k2}
\end{equation}

\begin{figure}[ht!]
    \plotone{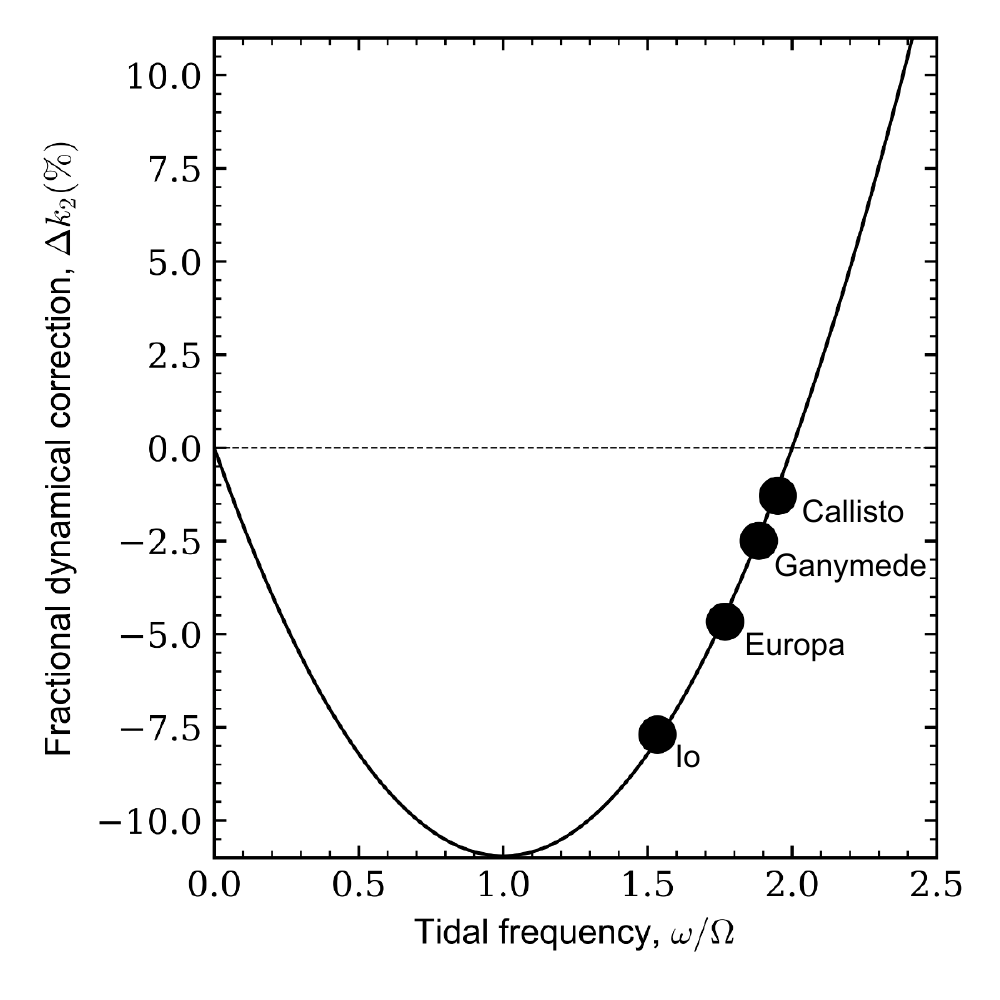}
    \caption{Fractional dynamical correction $\Delta k_2$ in a rotating uniform-density sphere including the Coriolis effect as a function of tidal frequency (\added{see equation} (\ref{eq3:k2})). \label{fig:k2}}
\end{figure}

Two effects contribute to the fractional dynamical correction: a negative contribution from the Coriolis effect $\propto 2\Omega\omega/\pi\mathcal{G}\bar{\rho}$ and a positive contribution from the dynamical amplification of f-modes $\propto \omega^2/\pi\mathcal{G}\bar{\rho}$.  The two contributions cancel each other at $2\Omega=\omega$, where the tide achieves hydrostatic equilibrium. Tides become hydrostatic not only when the planetary spin is phase-locked with the orbit of the satellite ($\Omega=\omega_s$), but also in planet-satellite systems where the central body is rotating at a rate orders of magnitude much faster than the orbit of the satellite (i.e., $\Omega\gg\omega_s$). As the frequency of the degree-2 f-mode approximately follows $\omega^2_0\sim g/R_p$, $\Delta k_2$ in equation (\ref{eq3:k2}) approximately becomes the positive fractional correction determined in Section~\ref{sec:norot} after setting $\Omega = 0$.

At the degree-2 Io-induced tidal frequency, the fractional dynamical correction corresponds to $\Delta k_2\approx-7.8\%$. The other Galilean satellites lead to a smaller $\Delta k_2$ because their tidal frequency falls closer to hydrostatic equilibrium (Fig. \ref{fig:k2}). A negative $\Delta k_2$ works in the direction required by the non-hydrostatic component identified by Juno in Jupiter's gravity field (Section~\ref{sec:junok2}).

\begin{figure}[ht!]
    \plotone{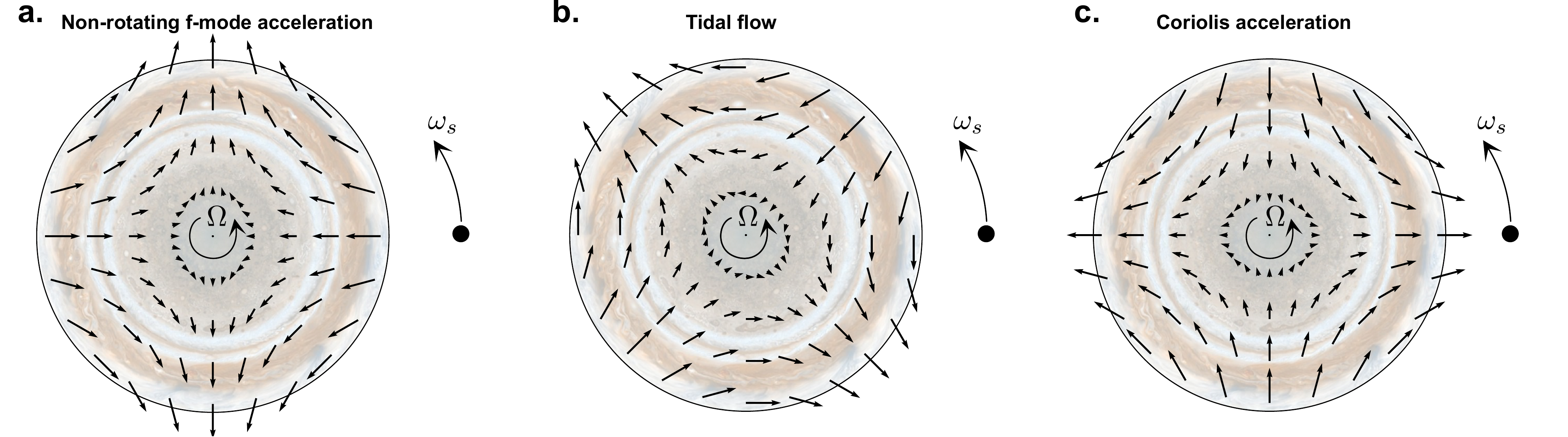}
    \caption{Degree-2 ($\ell=m=2$) tidal perturbations on a uniform-density sphere forced by the gravitational pull of a companion satellite: (a) the non-rotating f-mode acceleration $-\omega^2\bm{\xi}$, (b) tidal flow as shown in equation (\ref{eq3:vxy}), and (c) Coriolis acceleration $\bm{\Omega}\times\bm{v}$ according to the right-hand rule. \label{fig:planes}}
\end{figure}

  The direction of the flow provides an explanation for the negative sign of the fractional dynamical correction via the Coriolis acceleration. By definition, a uniform-density sphere has no density perturbations in its interior, thus produces an interior tidal gravitational potential that satisfies $\phi'\propto r^\ell Y^m_\ell \propto\phi^T$. We adopt equation (\ref{eq3:psi2}) as the degree-2 potential $\psi$ and obtain analytical solutions for the cartesian components of the resulting degree-2 tidal flow using equation (\ref{eq2:v}):
\begin{equation}
        \bm{v}_2 = -\frac{A\omega R_p}{g}\left(\hat{x}(ix+y)+\hat{y}(x-iy)\right)\textnormal{,}
\label{eq3:vxy}
\end{equation}
where $A$ is a constant depending on $\xi$ (Appendix~\ref{appendix:flow}). The degree-2 tidal flow purely exist in equatorial planes, showing no vertical component of motion (Fig.~\ref{fig:planes}b). 

The Coriolis acceleration plays a major role in setting the sign of the fractional dynamical correction for the Galilean satellites. Without Coriolis, the acceleration of non-rotating f-modes sustains a positive dynamical tidal displacement that follows $\xi^{dyn}\approx 5R_p\omega^2\xi^0/4g$. A $\xi^{dyn}>0$ increases the tidal gravitational field, which leads to a positive $\Delta k_2$. Conversely, as shown in equation (\ref{eq3:k2}), the fractional dynamical correction flips sign when Coriolis promotes $\xi^{dyn}<0$. A Coriolis term enters the momentum equation introducing an acceleration that competes with the acceleration of non-rotating f-modes, ultimately impacting $\xi^{dyn}$.  According to the right-hand rule, the Coriolis acceleration (i.e., $\bm{\Omega}\times\bm{v}$, Fig.~\ref{fig:planes}c) opposes the direction of the acceleration of non-rotating f-modes (i.e., $-\omega^2\bm{\xi}$, Fig.~\ref{fig:planes}a). The resulting gravitational field is smaller than the hydrostatic field if $\omega<2\Omega$, where the Coriolis acceleration beats the acceleration of non-rotating f-modes.

\subsubsection{The $n=1$ polytrope \label{sec:n1pol}}
In the following, we consider the more relevant case of a compressible planet that follows an $n=1$ polytropic equation of state (\ref{eq2:psi_n1}). In contrast to the localized tidal perturbation of a uniform-density sphere, a compressible body yields a tidally-induced density anomaly that arises from advection of the isodensity surfaces within the body. The resulting tidal gravitational potential is different in each case owing to differences in the tidally-perturbed density distribution obtained in a uniform-density sphere and a compressible body.

\begin{figure}[ht!]
    \plotone{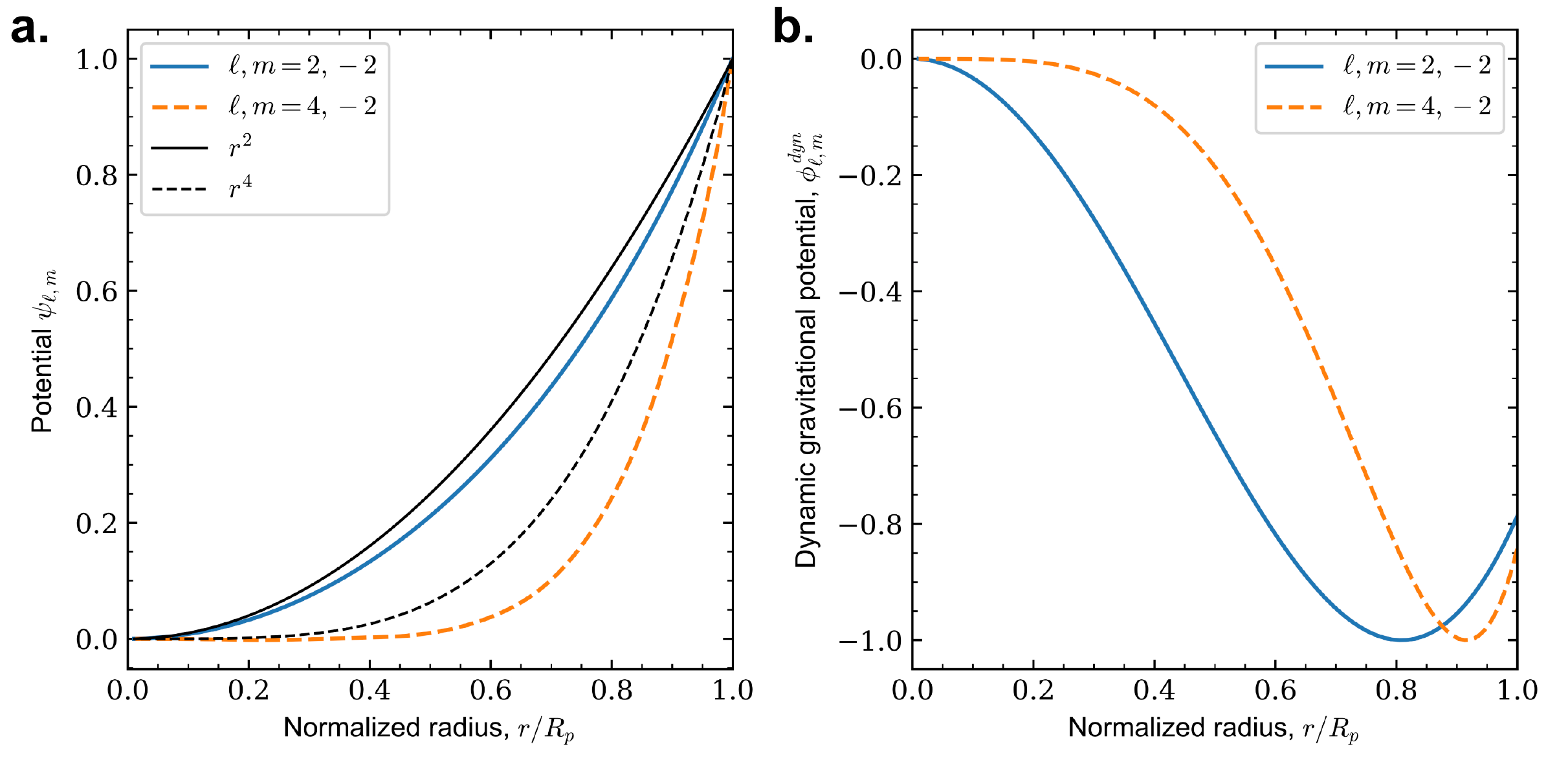}
    \caption{Radial functions in an $n=1$ polytrope (thick blue and orange curves) of the (a) potential $\psi$ and the (b) dynamical gravitational potential $\phi^{dyn}$. The thinner black curves in (a) represent the radial scaling of the potential $\psi$ in a uniform-density sphere. \label{fig:radialp}}
\end{figure}

Despite the aforementioned difference between models, the tidal flow remains similar so that dynamical tides motivate a negative correction to $k_2$ in each case. 
In an $n=1$ polytrope, the continuity equation (\ref{eq1:continuity}) tells us that the degree-2 radial component of the flow takes the form $v_r\propto j_2(kr)/j_1(kr)$ when the flow has small divergence, as it does.
Remarkably, the dominant contribution to the Taylor series expansion of $v_r$ is linear in $r$, even out to a large fraction of the planetary radius. 
In a uniform-density sphere, the potential $\psi$ is $\psi_2\propto r^2$, which leads to a tidal flow that  follows $v_2\propto\nabla\psi_2$; therefore, $v_r$ is also linear in $r$ in this model. 
As shown, the dominant contribution to $v_r$ scales with radius as $\propto r$, both in an $n=1$ polytrope and in a uniform-density sphere. In an $n=1$ polytrope, the dominant contribution to $v_r$ is curl-free and divergence-free and provides the $\psi_2\propto r^2$ part of the solution to the potential $\psi$ (Fig.~\ref{fig:radialp}a). Since the $n=1$ polytrope also contains terms where $\psi$ is higher order in $r$, it produces a flow with non-zero curl and non-zero divergence, causing $\psi_2$ to depart from $\psi_2\propto r^2$. Because high-order terms in $r$ are smaller than the dominant term, dynamical effects on $k_2$ in a uniform-density sphere are qualitatively similar to those in an $n=1$ polytrope. 

We compute the fractional dynamical correction to $k_2$ in a rotating polytrope following the same strategy used in Section~\ref{sec:norot2}. In opposition to the Coriolis-free polytrope, solving equation (\ref{eq2:psi_n1}) is technically challenging due to the $\ell-$coupling of the potential $\psi_{\ell,m}$ (e.g., mode mixing) promoted by the Coriolis effect. Mode mixing is also found in hydrostatic tides over a planet distorted by the effect of the centrifugal force \citep{wahl2017concentric}. The result of projecting equation (\ref{eq2:psi_n1}) into spherical harmonics is an infinite $\ell-$coupled set of ordinary differential equations for $\psi_{\ell,m }$ (Appendix \ref{sec:sh2}), similarly observed in the problem of dissipative dynamical tides \citep{ogilvielin2004tidal}. 
The Coriolis-promoted $\ell-$coupling comes from the sine and cosine in the spin rate of the planet ($\mathbf{\Omega}/\Omega=\hat{r}\cos\theta-\hat{\theta}\sin\theta$), which changes the degree of the spherical harmonics related to $\psi$. As a consequence, a given spherical harmonic from $\phi^0$ in the right-hand side in equation (\ref{eq2:psi_n1}) forces multiple spherical harmonics of the potential $\psi$ with different $\ell$.  

 Projected into spherical coordinates, the boundary condition (\ref{eq2:bcout}) at $r=R_p$ corresponds to:
\begin{equation}
    \partial_r\psi -\frac{2\Omega}{i\omega R_p}\partial_\varphi\psi - \frac{4\Omega^2}{\omega^2}\left(\cos^2\theta\partial_r\psi-\sin\theta\cos\theta\frac{\partial_\theta\psi}{R_p}\right)=\left(\frac{4\Omega^2-\omega^2}{g}\right)(\phi^0+\phi^T)\textnormal{.}
    \label{eq3:bcout}
\end{equation}
The outer boundary condition is also $\ell-$coupled after projected into spherical harmonics (Appendix~\ref{sec:sh2}).

At the center of the planet $r= r_0\to0$, we find the following scaling: $\nabla^2\phi^0\sim0$ and $j_0(kr_0)\sim \textnormal{constant}$. As a result, the tidal equation (\ref{eq2:psi_n1}) becomes the previously solved problem of the potential $\psi$ in a uniform-density sphere (\ref{eq3:uniform}) near the center. Requiring to be finite near the center and to satisfy equation (\ref{eq3:uniform}), the radial part of the potential $\psi$ follows $\psi_{\ell,m}\sim r^\ell$. The boundary condition for $\psi_{\ell,m}$ near the center corresponds to:
\begin{equation}
    \partial_r \psi_{\ell,m}- \frac{\ell}{r_0}\psi_{\ell,m} = 0\textnormal{.}
\end{equation}

The equation for the gravitational potential of dynamical tides $\phi^{dyn}$ remains unchanged compared to the Coriolis-free polytrope (Appendix \ref{sec:sh1}). The outer and inner boundary conditions for the gravitational potential generalize in degree as:
\begin{equation}
    \partial_r\phi^{dyn}_{\ell,m}  = -\left(\frac{(\ell+1)\phi^{dyn}_{\ell,m} + (2\ell+1)U_{\ell,m}}{R_p}\right)\textnormal{,}
\end{equation}
\begin{equation}
    \partial_r\phi^{dyn}_{\ell,m} -\frac{\ell}{r_0}\phi^{dyn}_{\ell,m} =0\textnormal{.}
\end{equation}

By projecting $\psi_{\ell,m}$ and $\phi_{\ell,m}$ into a series of $N$ Chebyshev polynomials oriented in the radial component (Appendix~\ref{sec:cheb}), we numerically solve (\ref{eq:coriolis_radial}) and (\ref{eq:phirad}) truncating the infinite series of $\ell-$coupled equations at an arbitrary $\ell=L_{max}$. We choose a truncation limit $L_{max}=50$ and the number of Chebyshev polynomials $N_{max}=100$ based on numerical evidence of convergence for $k_2$ and $k_{42}$.

 \begin{deluxetable*}{cccccccc}
\tablenum{2}
\tablecaption{Jupiter Love numbers.\label{tab:love}}
\tablewidth{0pt}
\tablehead{
\colhead{} & \colhead{Hydrostatic} & \colhead{Juno PJ17-$3\sigma$} & \colhead{3$\sigma$ fractional difference} &
\multicolumn{4}{c}{$\Delta k$ (rotating $n=1$ polytrope)}  \\
\colhead{Type} & \colhead{Number} & \colhead{Number} & \colhead{($\%$)} & \multicolumn{4}{c}{($\%$)} \\
 & & & &  \colhead{Io} & \colhead{Europa} & \colhead{Ganymede} & \colhead{Callisto} 
}
\decimalcolnumbers
\startdata
$k_2$   & 0.590  &  0.565 $\pm$ 0.018   & -7/-1 & \bf{-4} & -2 & -1 & -1 \\
$k_{42}$   & 1.743  &   1.289 $\pm$ 0.189    & -37/-15& \bf{+7}& +8 & +10& +12 \\
$k_{31}$ & 0.190 & 0.248 $\pm$ 0.046 & +6/+55 &\bf{+1}& +3 & +4& +5 \\
$k_{33}$ & 0.239 & 0.340 $\pm$ 0.116  & -6/+91 &\bf{ +2}& +5& +7 & +8 \\
$k_{44}$&0.135& 0.546 $\pm$ 0.406 & +4/+605 & \bf{+7}& +11& +13 & +15 \\
\enddata
\tablecomments{(2) The hydrostatic number is from \replaced{Wahl et al. (2016)}{\citet{wahl2020equilibrium}}. (3) The Juno PJ17-1$\sigma$ number is the satellite-independent number from \replaced{Notaro et al. 2019}{\citep{durante2020jupiter}}. (4) The 3$\sigma$ fractional difference represents the minimal/maximal  3$\sigma$ non-hydrostratic fractional correction required to explain Juno observations. The fractional dynamical correction in (5-8) is valid for an $n=1$ polytrope forced by the gravitational pull of the Galilean satellites. }
\end{deluxetable*}

We obtain $\Delta k_2 = -4.0\%$ at the degree-2 Io-induced tidal frequency (Table~\ref{tab:love}), which is of slightly lower amplitude than the estimate in a uniform-density sphere \added{and in agreement with the $k_2$ non-hydrostatic component observed by Juno at PJ17}. However different models, both the uniform-density sphere and the polytrope produce fractional dynamical corrections that fall within the order-of-magnitude estimate $\Delta k_2\sim\omega^2/4\pi\mathcal{G}\rho\sim 0.1$. As argued before, the dominant contribution to the potential $\psi$ follows the radial scaling $\psi_\ell\propto r^\ell$ (Fig.~\ref{fig:radialp}a). Ignoring the sign, the radial scaling of the dynamical gravitational potential $\phi^{dyn}$ (Fig.~\ref{fig:radialp}b) closely follows the shape of the hydrostatic gravitational potential (Fig.~\ref{fig:hsradial}).

\added{Due to the essentially circular and equatorial geometry of the Galilean orbits, the spherical harmonic $\ell=m=2$ dominates Jupiter's tidal gravitational field. Consequently, we concentrate in comparing $k_2$ Juno observation to our model prediction.} Of significantly higher uncertainty, the mid-mission Juno report of Love numbers at perijove 17 include other spherical harmonics in addition to $k_2$ (Table~\ref{tab:love}). Our polytropic \deleted{tidal} model predicts an \added{Io-induced tidal gravitational} field in a $3\sigma$-agreement with most Love numbers observed at PJ17, save for $k_{42}$ and $k_{31}$. 

\begin{figure}[ht!]
    \plotone{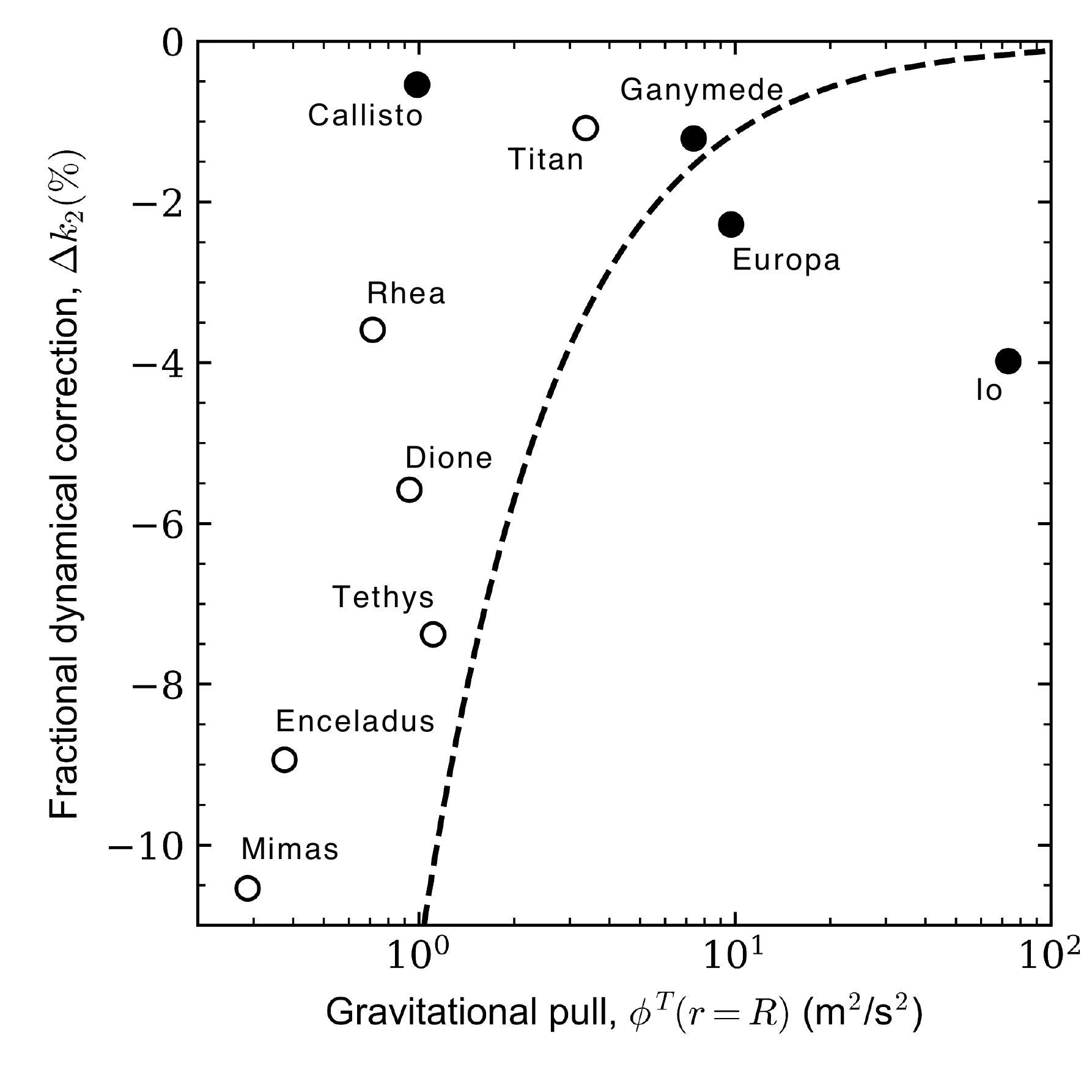}
    \caption{Conditions for the detection of dynamical tides evaluated for the Galilean satellites (black) and inner Saturn satellites (white). Satellites to the right of the dashed line have favorable conditions for a detection of dynamical tides assuming an uncertainty roughly similar to that of Io's $k_2$ on Jupiter at the end of Juno's extended mission. The fractional dynamical correction $\Delta k_2$ is for an $n=1$ polytrope. \label{fig:satellites}}
\end{figure}

\subsubsection{Detection of dynamical tides in systems other than Jupiter-Io.}
A detection of dynamical tides via direct measurement of the gravitational field will be challenging in bodies other than Jupiter (Fig. \ref{fig:satellites}). The $1\sigma$ uncertainty in the gravitational field of degree-2 Io tides is projected to be $\sigma_{J} \sim 6\cdot10^{-2}$ m$^2$/s$^2$ at the end of the proposed Juno extended mission (William Folkner, personal communication, April 8, 2020). The uncertainty in the measured tidal gravity field depends on the number and design of spacecraft orbits, the uncertainty in ephemerides, and instrumental capabilities. Assuming the uncertainty $\sigma_J$, we roughly estimate the gravitational pull required to produce a detectable dynamical component in the gravity field using:
\begin{equation}
    \phi^T(r=R)\gtrsim \frac{\sigma_{J}}{k_2|\Delta k_2|}\textnormal{.}
\end{equation}
Our calculation indicates that detecting dynamical tides in Saturn will require a mission with a more precise determination of the gravity field than that obtained by Juno (Fig. \ref{fig:satellites}). A $1\sigma$ detection of Europa-induced dynamical tides seems plausible at the end of Juno's extended mission, assuming that the factors determining the uncertainty in the gravity field remain similar to those of Io. \added{We calculate a model prediction for the satellite-dependent Jupiter Love number for all the Galilean satellites (Table~\ref{tab:love}). We obtain $k_2=0.578$ in the case of Europa, a prediction testable by the recently approved Juno extended mission.}

\section{Discussion}

\subsection{Future updates to Juno Love number observations.} 

The discrepancy between our predicted $k_{42}$, $k_{31}$, and the Juno-PJ17 observations may allude to several reasons: (1) a suggestion to revise the hydrostatic $\ell-$coupled $k_{\ell,m}$ in \replaced{Wahl et al. (2016)}{\cite{wahl2020equilibrium}}, (2) a failure of perturbation theory in our model when accounting for the $\ell-$coupled $k_{\ell,m}$, (3) other physical reasons; for example, tidal resonance with normal modes or the neglected correction from a dilute core. We strongly suggest a thorough analysis of these possibilities in future investigations. Ultimately, the perijove passes required to complete the scheduled Juno mission may change the still highly-uncertain numbers reported in Table~\ref{tab:love}. A recent revision to Juno observations at PJ29 (Daniele Durante, personal communication, November 18, 2020) suggest an agreement of our $k_{31}$ prediction with the revised satellite-independent $k_{31}= 0.234\pm0.016$ ($1\sigma$). The PJ29-revised $k_{42}=1.5\pm0.095$ ($1\sigma$) remains in disagreement with our $k_{42}$ prediction but the difference is much narrower than that attained at PJ17.

\added{A disagreement between our predicted $k_{\ell,m}$ and high-degree Juno observations does not impair the much more relevant agreement observed for $k_2$. Compared to the amplitude of the tidal gravitational potential related to $k_2$, the tidal gravitational potential related to $k_{42}$ represents an order-of-magnitude smaller contribution to the tidal gravitational field due to the factor $R_p/a\sim 1/6$ in $\phi^T_{\ell,m}$. In addition, whereas the predicted $k_2$ simply depends on the contribution from the Coriolis effect and the dynamical response of f-modes, the more complicated predicted $k_{42}$ additionally depends on the numerical solution of the $\ell-$coupled system of equations described in Section~\ref{sec:sh2}.}

Ignoring for the moment other possibilities \added{related to the $k_{42}$ discrepancy}, resonant tides have been previously invoked as a potential candidate to explain the current structure of the Laplace resonance in Saturn \citep{fuller2016resonance,lainey2020resonance}. As planets in the Solar System rotate far from break-up, there is no overlap between the frequencies of tides and f-modes in adiabatic, non-rotating planets. However, compositional gradients (g–modes) and rotation (inertial modes) introduce additional normal modes whose frequencies can become close to the tidal frequency, either by chance or planetary evolution. These hypothetically resonant tides could produce high dissipation rates; thus, a detectable imaginary part in the Love number that would consequently induce a significant change in the real part of the Love number. Equivalently, the high dissipation rate from an hypothetically resonant tide would cause a phase between the gravitational pull and the degree-2 tidal bulge. However, degree-2 tidal dissipation in Jupiter due to Io tides is modest \citep{lainey2009strong}. This argument does not necessarily apply for higher-degree tides ($\ell>2$) that have much smaller amplitudes and therefore whose phase shifts would be much harder to detect. \deleted{So far, we neither confirm or deny resonant tides that may be having an impact on $k_{42}$ or $k_{31}$.
A stronger conclusion on the possibility of tidal resonances observed in Juno data requires additional progress in the mission to reduce the uncertainty on $k_{42}$ and $k_{31}$, plus a thorough analysis of resonances with Jupiter interior models that include a compositional gradient. }

\added{We compare our Io-induced fractional dynamical corrections to the Juno-PJ17 satellite-independent observations. We justify the use of the satellite-independent $k_2$ uncertainty because our results indicate small variations in the Love number due to dynamical effects (Table~\ref{tab:love}), assuming the absence of degree-2 tidal resonances. In the hypothetical of an $\ell=4$, $m=2$ tidal resonance, the Love number $k_{42}$ would vary significantly among satellites. In such case, we would require to use a satellite-dependent uncertainty \citep{durante2020jupiter}, in which no a priori information is used at the time of inferring the Love number.  So far, we neither confirm or deny resonant tides that may be having an impact on $k_{42}$ or $k_{31}$. An improved version of the satellite-independent $k_2$ uncertainty could be obtained a priori assuming that the Love number increases $\sim4\%$ outward when comparing the inner to the outer satellites. A stronger conclusion on the possibility of tidal resonances observed in Juno data requires additional progress in the mission to reduce the uncertainty on $k_{42}$ and $k_{31}$, plus a thorough analysis of resonances with Jupiter interior models that include a compositional gradient. }

A tighter constraint on the \replaced{planetary}{satellite-dependent} $k_2$ from satellites other than Io \replaced{may confirm}{will test} the prediction of our model of \added{satellite-dependent} dynamical tides\deleted{or motivate the introduction of additional corrections}. Despite the relatively large fractional dynamical correction obtained for the inner Saturn satellites (e.g. Mimas or Enceladus), their small mass leads to an overall small tidal disturbance that is difficult to detect in the gravity field. From all Jupiter and Saturn satellites, only Europa elevates a short-term prospect of obtaining a new detection of dynamical tides via Juno's extended mission (Fig.~\ref{fig:satellites}). A detection of dynamical tides due to other satellites will require an uncertainty on $k_2$ significantly lower than that produced by Juno.

\subsection{Other potential contributions to $k_2$}
\subsubsection{Free-oscillating normal modes}
Free oscillations of normal modes cannot explain the bulk of the non-hydrostatic Juno detection discussed here. The small gravitational field of tides becomes resolvable by Juno in part because the phase of the signal is well known. The unknown phase of non-resonant free oscillations departs from the phase of the satellite used in determining $k_2$. Even if freely-oscillating normal modes were detected in Jupiter as they were in Saturn (Iess et al, 2019), the anticipated high frequency of their gravity field would render them irrelevant to the tidal problem. In order that free oscillations play a role in the observed gravity, they require to avoid a rapid decay after becoming excited (i.e., a very high $Q$). It is not known whether free oscillations persist over multiple Juno perijove passes. 

\subsubsection{Jupiter's rheology}
A central viscoelastic region in Jupiter's interior could potentially reduce $k_2$ below the hydrostatic number; however, evidence suggest that such possibility is unlikely. Viscoelastic deformation of a body produces a $k_2$ between the purely elastic and the hydrostatic number; a model that helps to explain Titan's observed $k_2$ \citep{iess2012tides}. Assuming that a traditional core in Jupiter exists, the core radius should remain small (i.e., $\sim0.15R_J$) to satisfy the constraint on the total abundance of heavy elements and the super-solar enrichment of the envelope \replaced{Wahl et al. (2016)}{\citep{wahl2020equilibrium}}. At this core radius, the tidal deformation of the core does not contribute to  Re($k_2$) \citep{storch2014viscoelastic}. Whether rigid, elastic, or viscoelastic, a small traditional core produces a small effect on Re($k_2$) due to the added heavy elements, already included in the hydrostatic number \replaced{Wahl et al. (2016)}{\citep{wahl2020equilibrium}}. Beyond the possibility of a viscoelastic traditional core, the hydrogen-rich envelope most likely behaves as an inviscid fluid.
The kinematic viscosity of the fluid external to the core needs to reach $\nu\sim 10^{11}$ m$^2$/s in order for its viscosity to become relevant at tidal timescales (i.e, $\omega\sim \nu/R^2_J $). Such kinematic viscosity exceeds by $\sim 17$ orders of magnitude realistic estimates of the hydrogen-dominated fluid viscosity \citep{stevenson1977dynamics}. A similar argument applies to a dilute core, which most likely consist of a mixture dominated by hydrogen, either by atomic number and even probably by mass.

\subsubsection{The dynamical contribution of a traditional core}
A traditional core blocks the tidal flow from extending to the center of the planet by forcing a zero-flow boundary condition at the core radius. As shown earlier, the radial tidal flow sets the amplitude of the tidal gravitational potential and roughly scales with distance from the center following $v_r\propto r$ in an $n=1$ polytrope.  Consequently, the tidal flow is nearly zero in the area where a traditional core would exist, minimizing a potential effect of the traditional core on the fractional dynamical correction. A thorough quantification of $\Delta k_2$ in a model with a traditional core that blocks the flow requires further investigation. \added{We expect an effect going from negligible to small (i.e., less than $+1\%$ applied to the current estimate in Table~\ref{tab:love}) given the limits to traditional core size imposed by the constrained total abundance of heavy elements.}

\subsubsection{A dilute core}
A dilute core may promote an additional departure of the tidal response from the hydrostatic tide to that caused by dynamical tides. The hydrostatic tide in $k_2$ provides the same information about the planet than $J_2$ \citep{hubbard1984planetary}. In the presence of a dilute core, the gravity produced by tides fundamentally differ from the $J_{2\ell}$ coefficients due to the different timescales associated to tidal perturbations and evolution of the rotation rate. Tidal timescales are short compared to the timescale required for the tidal perturbation to equilibrate with the environment by either heat transport or compositional evolution. By contrast, the timescale at which the rotation rate evolves is so long that the planet adjust any perturbation caused by the centrifugal effect. Tidal displacements remain roughly adiabatic whereas displacements induced by changes in the rotation rate reach thermodynamic equilibrium. A fluid parcel in the proximity of the dilute core responds differently depending on the timescale of the perturbation; only an adiabatic perturbation leads to changes in the buoyancy of the fluid parcel, causing a wave-like oscillation known as static stability. Consequently, the dilute core produces a signature in the tidal response of the planet not registered by $J_2$. We address the tidal effects of a dilute core in a subsequent investigation.

\section{Conclusions}
Our tidal models suggest that the gravity field observed by Juno captured the dynamical tidal response of Jupiter to the gravitational pull of the Galilean satellites. We show that two effects contribute to the dynamical gravity field of tides in Jupiter: the dynamical response of f-modes and the Coriolis acceleration. When the Coriolis effect is ignored, tides closely follow the dynamical response of f-modes modeled as a forced harmonic oscillator. In ignoring the Coriolis effect, dynamical amplification in a harmonic oscillator accounts for the dynamical response of f-modes in the planet's interior, forced by the gravitational pull of the companion satellites. As the tidal frequency is lower than the f-mode oscillation frequency, the dynamical response of f-modes amplifies the gravity field of the hydrostatic tide. Motivated by Jupiter's fast rotation, we show that the Coriolis effect leads to a significant additional contribution to the dynamical tide. When the Coriolis effect is included in our tidal models, we show that the Coriolis acceleration produces a competing effect of opposite sign compared to the dynamical response of f-modes. When both dynamical effects are considered together, dynamical effects reduce the Love number $k_2$ below the hydrostatic number if $\omega < 2\Omega$ and amplify it otherwise. Following our theoretical prediction, dynamical effects lead to a negative correction to Jupiter's hydrostatic Love number $k_2$ in the case of the Galilean satellites, which degree-2 tidal frequency is $\omega<2\Omega$. The fractional dynamical correction for the Jupiter-Io system is \replaced{$\Delta k_2 = -4.0\%$}{$\Delta k_2 = -4\%$}. Our analysis provides an explanation for the recently observed non-hydrostatic component in the gravity field of Jupiter tides obtained by the Juno mission.

In conclusion, our analysis proposes that the Juno non-hydrostatic detection is the first unambiguous measurement of the gravitational effects of dynamical tides in a gas giant. \added{Our conclusion depends on the assumption that the degree-2 tidal frequency of the Galilean satellites is far from resonance with Jupiter's normal modes.} In a subsexquent investigation, we utilize the results reported here to infer the extension and static stability of Jupiter's dilute core from $k_2$. The uncertainty expected in the observed $k_2$ at the end of the mission exceeds the uncertainty achieved by our model, suggesting that a more detailed tidal model will be required in the future to fully exploit the information contained in the data provided by Juno. 


\acknowledgments
We thank the support of NASA's Juno mission. B.I. thanks Erin Burkett for her comments. We acknowledge the constructive comments from two anonymous referees. 

%

\vspace{5mm}


\software{Matplotlib \citep{hunter2007matplotlib}
          }



\appendix

\section{Hydrostatic tides in an index-one polytrope \label{sec:hs}}

In hydrostatic tides, the tidal frequency becomes $\omega \approx 0$ and the tidal flow is slow enough to set $\bm{v}\approx 0$. After projecting $\phi^0$ into spherical harmonics by setting a solution in the form:
\begin{equation}
    \phi^0= \sum_{\ell,m}\phi_{\ell,m}^0(x)Y_m^\ell(\theta,\varphi)e^{-i\omega t}\textnormal{,}
\end{equation}
the radial part of $\phi^0$ at a given harmonic that satisfies equation (\ref{eq:8}) follows:

\begin{equation}
   \partial_{x,x} \phi^0_\ell + \frac{2}{x} \partial_x\phi_\ell^0 + \left( 1 -  \frac{\ell(\ell+1)}{x^2}\right) \phi_\ell^0  = -\left(\frac{x}{\pi}\right)^\ell\textnormal{,}
   \label{eq:static}
\end{equation}
where $Y_\ell^m$ are normalized spherical harmonics defined by:
\begin{equation}
    Y_\ell^m(\theta,\varphi) =  \sqrt{\frac{(2\ell+1)}{4\pi}\frac{(\ell-m)!}{(\ell+m)!}}\mathcal{P}_\ell^m(\cos\theta)e^{im\varphi}\textnormal{,}
\end{equation}
and $\mathcal{P}_\ell^m$ Associated Legendre polynomials corresponding to:
\begin{equation}
\mathcal{P}_\ell^m(\mu) = \frac{(-1)^m}{2^\ell l!}(1-\mu^2)^{m/2}\frac{d^{\ell+m}}{d\mu^{\ell+m}}(\mu^2-1)^\ell\textnormal{.}
\end{equation}
The normalized radial coordinate follows $x = kr$, which leads to a planet with radius $\pi$. Note that equation (\ref{eq:static}) is non-dimensional and should be scaled by the factor:
\begin{equation}
    U_{\ell,m} = \left(\frac{\mathcal{G}m_s}{a}\right) \left(\frac{R_p}{a}\right)^\ell\left( \frac{4\pi (\ell-m)!}{(2\ell + 1)(\ell+m)!}\right)^{1/2} \mathcal{P}_l^{m}(0)\textnormal{.}
\end{equation}
The order $m$ does not appear in equation (\ref{eq:static}), indicating a degeneracy on $m$ of the hydrostatic tide. As $x^\ell Y_\ell^m$ is a solution to Laplace's equation (i.e., $\nabla^2(x^\ell Y_\ell^m) = 0$), a complete solution to equation (\ref{eq:static}) is:

\begin{equation}
    \phi^0_\ell = A j_\ell(x) + B n_\ell(x) - \left(\frac{x}{\pi}\right)^\ell\textnormal{.}
\end{equation}
We require $\phi^0$ to be finite at the center of the planet and thus set $B=0$. According to the outer boundary condition, we set a external gravitational potential $\Phi_\ell^0(x)$ that extends outward from the planet and matches the internal tidal potential at the planetary radius as:
\begin{equation}
    \Phi_\ell^0 (x)= \left(\frac{\pi}{x}\right)^{\ell+1}\phi_\ell^0(\pi) = \left(\frac{\pi}{x}\right)^{\ell+1} (A_\ell j_\ell(\pi) - 1)\textnormal{.}
\end{equation}
The continuity of the gradients of the internal and external potentials at the surface of the planet sets the constant $A_\ell$ to:
\begin{equation}
    A_\ell = \frac{2\ell +1}{\pi j_{\ell-1}(\pi)}\textnormal{.}
\end{equation}
Consequently, the gravitational potential of hydrostatic tides at degree $\ell$ is:
\begin{equation}
    \phi^0_\ell = \left(\frac{2\ell +1}{\pi}\right) \frac{j_\ell(x)}{j_{\ell-1}(\pi)}-\left(\frac{x}{\pi}\right)^\ell \textnormal{,}
    \label{eqa:phi0}
\end{equation}
and the hydrostatic Love number follows:
\begin{equation}
    k_{\ell} = \left(\frac{2\ell +1}{\pi}\right) \frac{j_\ell(\pi)}{j_{\ell-1}(\pi)} - 1 \textnormal{.}
\end{equation}

\section{Projection of the dynamical tide equations into spherical harmonics \label{sec:sh}}

Here we project into spherical harmonics the equation for the potential $\psi$ in a non-rotating (\ref{eq3:approximation}) and rotating (\ref{eq2:psi_n1}) $n=1$ polytrope. The equation for the gravitational potential of dynamical tides is equation (\ref{eq2:phi_n1}), forced by a different potential $\psi$ depending on rotation. We evaluate solutions in the form:
\begin{equation}
        \psi = \sum_{\ell,m}\psi_{\ell,m}(x) Y_\ell^m(\theta,\varphi)e^{-i\omega t}\textnormal{,}
\end{equation}
\begin{equation}
    \phi^{dyn} = \sum_{\ell,m}\phi^{dyn}_{\ell,m}(x) Y_\ell^m(\theta,\varphi)e^{-i\omega t}\textnormal{.}
\end{equation}
 In the following, we conveniently drop the time dependent part $e^{i\omega t}$ out of our derivation. Notice that we normalize the radial coordinate following $x=kr$, leading to a body of normalized radius $\pi=kR_p$.

\subsection{The Coriolis-free $n=1$ polytrope \label{sec:sh1}}

We project into spherical harmonics the potential $\psi$ (\ref{eq3:approximation}) and the dynamical gravitational potential (\ref{eq2:phi_n1}) of the non-rotating polytrope:
\begin{equation}
    j_0(x)\left(\partial_{x,x} (\psi_{\ell,m}) + \left(\frac{2}{x}-\frac{j_1(x)}{j_0(x)}\right) \partial_x(\psi_{\ell,m}) + \left( 1 -  \frac{\ell(\ell+1)}{x^2}\right) \psi_{\ell,m} \right)  = \left(\frac{2\ell +1}{\pi j_{\ell-1}(\pi)}\right) \frac{\omega^2 j_\ell(x)}{4\pi\mathcal{G}\rho_c} \textnormal{,}
\end{equation}
\begin{equation}
   \partial_{x,x} \left(\phi^{dyn}_{\ell,m}\right) + \frac{2}{x} \partial_x\left(\phi_{\ell,m}^{dyn}\right) + \left( 1 -  \frac{\ell(\ell+1)}{x^2}\right) \phi_{\ell,m}^{dyn}  = \psi_{\ell,m}\textnormal{.}
   \label{eq:phirad}
\end{equation}

\subsection{The $n=1$ polytrope \label{sec:sh2}}
 Relative to the left-hand side in equation (\ref{eq2:psi_n1}), the projection of the first, second, and third terms, respectively follow:
\begin{equation}
   \nabla\cdot\left(j_0 \nabla\psi_{\ell,m}\right) = j_0\left( \partial_{x,x} + \left(\frac{2}{x}- \frac{j_1}{j_0}\right)\partial_x  - \frac{\ell(\ell+1)}{x^2} \right) \psi_{\ell,m}Y_\ell^m\textnormal{,}
\end{equation}
\begin{equation}
    \frac{2}{i\omega}\nabla\cdot(j_0\mathbf{\Omega}\times\nabla\psi_{lm})= \frac{2m\Omega j_1}{\omega x}\psi_{\ell,m}Y_\ell^m \textnormal{,}
\end{equation}
\begin{eqnarray}
  \label{eqap:third}
  -\frac{4}{\omega^2}  \nabla\cdot\left(j_0\mathbf{\Omega}(\mathbf{\Omega}\cdot\nabla\psi_{\ell,m})\right) = -\frac{4\Omega^2}{\omega^2} \left(\left(j_0\partial_{x,x}-\left(j_1+\frac{j_0}{x}\right)\partial_x  \right)\psi_{\ell,m}Y_\ell^m\cos^2\theta\right. \nonumber\\+ \frac{j_0}{x}\partial_x\psi_{\ell,m}Y_\ell^m \left. +\left(\frac{2j_0}{x^2}+\frac{j_1}{x}-\frac{2j_0}{x}\partial_x\right)\psi_{\ell,m}\cos\theta\sin\theta\partial_\theta Y_\ell^m  + \frac{j_0}{x^2}\psi_{\ell,m} \sin^2\theta\partial_{\theta,\theta} Y_\ell^m\right)\textnormal{.}
\end{eqnarray}
The multiplication of spherical harmonics with trigonometric functions expresses a physical statement about the coupling effect that Coriolis produces in the tidal gravitational response of a rotating body. The partial derivatives in the spherical harmonics indicate changes in quantum numbers described in the following differential relations \citep{lockitch1999r}:
\begin{equation}
    \sin\theta\partial_\theta Y_\ell^m   = \ell Q_{\ell+1}Y_{\ell+1}^m - (\ell+1)Q_\ell Y_{\ell-1}^m\textnormal{,}
\end{equation}
\begin{equation}
   \cos\theta Y_\ell^m   = Q_{\ell+1}Y_{\ell+1}^m+Q_\ell Y_{\ell-1}^m\textnormal{,}
\end{equation}
where
\begin{equation}
    Q_\ell = \left(\frac{\ell^2 - m^2}{4\ell^2 -1}\right)^{1/2}\textnormal{.}
\end{equation}
 Combining the previous differential relations, we arrive to expressions for each of the angular terms in equation (\ref{eqap:third}):
\begin{equation}
    Y_\ell^m\cos^2\theta = Q_{\ell-1} Q_\ell Y_{\ell-2}^m + (Q_\ell^2 + Q_{\ell+1}^2) Y_\ell^m + 
 Q_{\ell+1} Q_{\ell+2} Y_{\ell+2}^m\textnormal{,}
\end{equation} 
\begin{equation}
    \cos\theta\sin\theta\partial_\theta Y_\ell^m = -(\ell+1) Q_{\ell-1} Q_{\ell} Y_{\ell-2}^m - ((\ell+1) Q_\ell^2 - \ell Q_{\ell+1}^2) Y_\ell^m + \ell Q_{\ell+1} Q_{\ell+2} Y_{\ell+2}^m\textnormal{,}
\end{equation} 
\begin{equation}
    \sin^2\theta\partial_{\theta,\theta} Y_\ell^m = (\ell+1)^2 Q_{\ell-1} Q_\ell Y_{\ell-2}^m + ((2 + \ell - \ell^2) Q_\ell^2 - \ell (\ell+3) Q_{\ell+1}^2) Y_\ell^m + \ell^2 Q_{\ell+1} Q_{\ell+2} Y_{\ell+2}^m\textnormal{.}
\end{equation}
After grouping terms with the same spherical harmonic, equation (\ref{eq2:psi_n1}) becomes an infinite set of $\ell$-coupled radial equations following the structure of a Sturn-Liouville problem:
\begin{eqnarray}
\label{eq:coriolis_radial}
    \left((P_{\ell,m}^{(1)}\partial_{x,x} + Q_{\ell,m}^{(1)}\partial_x + R_{\ell,m}^{(1)}\right)\psi_{\ell,m} + \left(P_{\ell,m}^{(2)}\partial_{x,x} + Q_{\ell,m}^{(2)}\partial_x + R_{\ell,m}^{(2)}\right)\psi_{l+2,m} \nonumber\\+ \left(P_{\ell,m}^{(0)}\partial_{x,x} + Q_{\ell,m}^{(0)}\partial_x + R_{\ell,m}^{(0)}\right)\psi_{l-2,m} \nonumber\\ = U_{\ell,m} \left(\frac{\omega^2-4\Omega^2}{4\pi\mathcal{G} \rho_c}\right)\left(\frac{2\ell +1}{\pi}\right) \frac{j_\ell(x)}{j_{\ell-1}(\pi)}\textnormal{.}
\end{eqnarray}
The nine radial coefficients correspond to:
\begin{equation}
    P_{\ell,m}^{(0)} =- \frac{4\Omega^2}{\omega^2} j_0Q_{\ell-1}Q_\ell \textnormal{,}
\end{equation}
\begin{equation}
    P_{\ell,m}^{(1)} = j_0 \left( 1- \frac{4\Omega^2}{\omega^2}(Q_\ell^2+Q_{\ell+1}^2)\right)\textnormal{,}
\end{equation}
\begin{equation}
    P_{\ell,m}^{(2)} =- \frac{4\Omega^2}{\omega^2}j_0Q_{\ell+1}Q_{\ell+2}\textnormal{,}
\end{equation}
\begin{equation}
    Q_{\ell,m}^{(0)} = \frac{4\Omega^2}{\omega^2}\left((2\ell-3)\frac{j_0}{x}+j_1\right)Q_{\ell-1}Q_\ell\textnormal{,}
\end{equation}
\begin{equation}
    Q_{\ell,m}^{(1)} = \frac{2j_0}{x} -j_1 - \frac{4\Omega^2}{\omega^2}\left(\frac{j_0}{x}\left( 1+(2\ell+1) (Q_\ell^2-Q_{\ell+1}^2)\right) -j_1 (Q_\ell^2+Q_{\ell+1}^2)\right)\textnormal{,}
\end{equation}
\begin{equation}
     Q_{\ell,m}^{(2)} = -\frac{4\Omega^2}{\omega^2}\left((2\ell-3)\frac{j_0}{x}-j_1\right)Q_{\ell+1}Q_{\ell+2}\textnormal{,}
\end{equation}
\begin{equation}
    R_{\ell,m}^{(0)} = -\frac{4\Omega^2}{\omega^2}(\ell-2)\left(\ell\frac{j_0}{x^2}+\frac{j_1}{x}\right)Q_{\ell-1}Q_\ell\textnormal{,}
\end{equation}
\begin{equation}
    R_{\ell,m}^{(1)} = -\frac{l(l+1)j_0}{x^2} + \frac{2m\Omega j_1}{\omega x} + \frac{4\Omega^2}{\omega^2}\left( \frac{j_1}{x}((\ell+1)Q_\ell^2-\ell Q_{\ell+1}^2) + \frac{j_0}{x^2}\ell(\ell+1)(Q_\ell^2+Q_{\ell+1}^2) \right)\textnormal{,}
\end{equation}
\begin{equation}
    R_{\ell,m}^{(2)} = -\frac{4\Omega^2}{\omega^2}\left((\ell(\ell+4)+11)\frac{j_0}{x^2}-(\ell-1)\frac{j_1}{x}\right)Q_{\ell+1}Q_{\ell+2}\textnormal{.}
\end{equation}

The projection into spherical harmonics of the boundary condition (\ref{eq3:bcout}) leads to:
\begin{eqnarray}
      Y_\ell^m\left(\partial_x\psi_{\ell,m} -\frac{2m\Omega}{\omega \pi}\psi_{\ell,m}\right) &-& \frac{4\Omega^2}{\omega^2}\left(Y_\ell^m\cos^2\theta\partial_x\psi_{\ell,m}-\frac{\psi_{\ell,m}}{\pi}\sin\theta\cos\theta\partial_\theta Y_\ell^m\right)\nonumber\\&=&U_{\ell,m}\left(\frac{4\Omega^2-\omega^2}{g}\right)\left(\frac{2\ell +1}{\pi}\right) \frac{j_\ell(x)}{j_{\ell-1}(\pi)}Y_\ell^m\textnormal{.}
\end{eqnarray}
The previous differential relations still apply to deal with coupled spherical harmonics in the boundary condition. Grouping terms for each spherical harmonic $Y_\ell^m$, we reach an $\ell-$coupled boundary condition with the structure:
\begin{equation}
    \sum_{j=0}^2\left(\hat{Q}_{\ell,m}^{(j)}\partial_x + \hat{R}_{\ell,m}^{(j)}\right) \psi_{\ell +2j-2}^m= U_{\ell,m}\left(\frac{4\Omega^2-\omega^2}{g}\right)\left(\frac{2\ell +1}{\pi}\right) \frac{j_\ell(\pi)}{j_{\ell-1}(\pi)}\textnormal{.}
\end{equation}
The six radial coefficients correspond to:
\begin{equation}
    \hat{Q}_{\ell,m}^{(0)} = -\frac{4\Omega^2}{\omega^2} Q_{\ell-1} Q_\ell\textnormal{,}
\end{equation}
\begin{equation}
    \hat{Q}_{\ell,m}^{(1)} = 1-\frac{4\Omega^2}{\omega^2}\left(Q_\ell^2+Q^2_{\ell+1}\right)\textnormal{,}
\end{equation}
\begin{equation}
    \hat{Q}_{\ell,m}^{(2)} = -\frac{4\Omega^2}{\omega^2}Q_{\ell+1}Q_{\ell+2}\textnormal{,}
\end{equation}
\begin{equation}
    \hat{R}_{\ell,m}^{(0)} = \frac{4\Omega^2}{\pi\omega^2}(\ell-2)Q_{\ell-1}Q_\ell\textnormal{,}
\end{equation}
\begin{equation}
    \hat{R}_{\ell,m}^{(1)} = -\frac{2m\Omega}{\pi\omega}-\frac{4\Omega^2}{\pi\omega^2}\left((\ell+1)Q_\ell^2-\ell Q_{\ell+1}^2\right)\textnormal{,}
\end{equation}
\begin{equation}
    \hat{R}_{\ell,m}^{(2)} = -\frac{4\Omega^2}{\pi\omega^2}(\ell-1)Q_{\ell+1}Q_{\ell+2}\textnormal{.}
\end{equation}

\section{Chebyshev pseudo-spectral method \label{sec:cheb}}

We solve the Sturn-Liouville differential problem \citep{boyd2001chebyshev} defined by:

\begin{equation}
    p(r) u''(r) + q(r) u'(r) + r(r) u(r) = f(r)\textnormal{,}
    \label{eq:SL}
\end{equation}
and constrained to the boundary conditions:
\begin{equation}
    \alpha_0 u'(a) + \alpha_1 u(a) = \alpha_2\textnormal{,}
\end{equation}
\begin{equation}
    \beta_0 u'(b) + \beta_1 u(b) = \beta_2\textnormal{,}
\end{equation}
where $a$ and $b$ are the two ends of a boundary value problem. We shift the domain of equation (\ref{eq:SL}) from $r \in [a,b]$ to the domain of Chebyshev polynomials $\mu\in[-1,1]$ and seek for a solution that is a truncated sum of an infinite Chebyshev series:
\begin{equation}
    u(\mu) \approx \sum_{n=0}^{N_{max}} a_n T_n(\mu)\textnormal{,}
\end{equation}
with the Chebyshev polynomials defined by:
\begin{equation}
    T_n(\mu) = \cos(nt)\textnormal{,}
\end{equation}
and $t = \arccos(\mu)$. Our objective is to obtain the coefficients $a_n$ by solving a linear inverse problem:
\begin{equation}
    La = f\textnormal{.}
\end{equation}
The square matrix $L$ and the vector $f$ come from the evaluation of equation (\ref{eq:SL}) into Gauss-Lobatto collocation points defined by:
\begin{equation}
    \mu_i = \cos\left(\frac{\pi i}{N-1}\right)\textnormal{,}\hspace{2cm} i=1,2,\dots,N-1\textnormal{,}
\end{equation}
plus the constraints from boundary conditions. The partial derivatives in equation (\ref{eq:SL}) assume the analytical form:
\begin{equation}
    \frac{\partial T_n(\mu)}{\partial \mu} = n \frac{\sin(nt)}{\sin(t)}\textnormal{,}
\end{equation}
\begin{equation}
    \frac{\partial^2 T_n(\mu)}{\partial \mu^2} = -n^2 \frac{\cos(nt)}{\sin^2(t)} + \left(\frac{n\cos(t)}{\sin^3(t)}\right)\sin(nt)\textnormal{.}
\end{equation}

\section{Tidal flow in a uniform-density sphere \label{appendix:flow}}
We calculate the tidal flow from projecting equation (\ref{eq2:v}) into cartesian coordinates:
\begin{equation}
    \bm{v} = -\left(\frac{i\omega}{4\Omega^2-\omega^2}\right)\left( \hat{x}\left(\partial_x +\frac{2i\Omega}{\omega}\partial_y\right)+\hat{y}\left(\partial_y-\frac{2i\Omega}{\omega}\partial_x\right)+\hat{z}\left(\frac{4\Omega^2}{\omega^2}-1\right)\partial_z\right)\psi\textnormal{.}
    \label{eq:vcart}
\end{equation}
The potential $\psi$ depends on the potential of the gravitational pull (\ref{eq1:forcing2}) and the tidal potential internal to the thin shell disturbed by tides. Analogously as we did for the external potential, we obtain the  tidal gravitational potential (i.e., $r<R_p$) from integration throughout the volume:
\begin{equation}
    \phi_2'= \frac{3}{5}\left(\frac{r}{R_p}\right)^2g\xi_2\textnormal{,}
\end{equation}
which leads to:
\begin{equation}
    \psi_2 = \frac{ R_p\omega(2\Omega-\omega)}{2g}\tilde{\phi_2}'= \frac{ AR_p\omega(2\Omega-\omega)}{2g}(x-iy)^2\textnormal{.}
    \label{eq:psi2ap}
\end{equation}
The constant $A$ comes from the numerical factor of the relevant potentials, corresponding to:
\begin{equation}
    A = \frac{3}{16}\frac{\mathcal{G}m_s}{a^3} + \frac{3}{20}\sqrt{\frac{15}{2\pi}}\frac{g\xi_2}{R_p^2}\textnormal{.}
\end{equation}

As shown in equation (\ref{eq:psi2ap}), $\psi_2$ is independent of $z$, leading to a 2-D tidal flow in equatorial planes. Replacing equation (\ref{eq:psi2ap}) into equation (\ref{eq:vcart}), the degree-2 tidal flow becomes:
\begin{equation}
        \bm{v}_2 = -\frac{A\omega R_p}{g}\left(\hat{x}(ix+y)+\hat{y}(x-iy)\right)\textnormal{.}
\end{equation}





\listofchanges

\end{document}